\documentclass[sn-mathphys,Numbered]{sn-jnl}

\usepackage{graphicx}%
\usepackage{multirow}%
\usepackage{amsmath,amssymb,amsfonts}%
\usepackage{amsthm}%
\usepackage{mathrsfs}%
\usepackage[title]{appendix}%
\usepackage{xcolor}%
\usepackage{textcomp}%
\usepackage{manyfoot}%
\usepackage{booktabs}%
\usepackage{algorithm}%
\usepackage{algorithmicx}%
\usepackage{algpseudocode}%
\usepackage{listings}%
\usepackage{wrapfig}
\usepackage{lineno}
\usepackage[normalem]{ulem}

\theoremstyle{thmstyleone}%
\theoremstyle{thmstyletwo}%

\theoremstyle{thmstylethree}%

\begin{document}

\title[[Article Title]{Clouds dissipate quickly during solar eclipses as the land surface cools}


\author*[1,2]{\fnm{Victor J. H. Trees}}\email{v.j.h.trees@tudelft.nl}

\author[2]{\fnm{Stephan R. de Roode}}

\author[1,3]{\fnm{Job I. Wiltink}}

\author[1]{\fnm{Jan Fokke Meirink}}

\author[1]{\fnm{Ping Wang}}

\author[1]{\fnm{Piet Stammes}}

\author[2]{\fnm{A. Pier Siebesma}}


\affil[1]{\orgdiv{R\&D Satellite Observations}, \orgname{Royal Netherlands Meteorological Institute (KNMI)}, \city{de Bilt}, \country{The Netherlands}}

\affil[2]{\orgdiv{Geoscience and Remote Sensing}, \orgname{Delft University of Technology}, \city{Delft}, \country{The Netherlands}}

\affil[3]{\orgdiv{Meteorology and Air Quality Group}, \orgname{Wageningen University \& Research}, \city{Wageningen}, \country{The Netherlands}}




\maketitle

\section*{Abstract}

\textbf{Clouds affected by solar eclipses could influence the reflection of sunlight back into space and might change local precipitation patterns. Satellite cloud retrievals have so far not taken into account the lunar shadow, hindering a reliable spaceborne assessment of the eclipse-induced cloud evolution. Here we use satellite cloud measurements during three solar eclipses between 2005 and 2016 that have been corrected for the partial lunar shadow together with large-eddy simulations to analyze the eclipse-induced cloud evolution. Our corrected data reveal that, over cooling land surfaces, shallow cumulus clouds start to disappear at very small solar obscurations ($\sim$15\%). Our simulations explain that the cloud response was delayed and was initiated at even smaller solar obscurations. We demonstrate that neglecting the disappearance of clouds during a solar eclipse could lead to a considerable overestimation of the eclipse-related reduction of net incoming solar radiation. These findings should spur cloud model simulations of the direct consequences of sunlight-intercepting geoengineering proposals, for which our results serve as a unique benchmark.}

\section*{Introduction}

Blocking part of the solar radiation incident on the Earth's (lower) atmosphere and surface is one of the proposed strategies to counteract the current and future global temperature rise, which may be inevitable if climate change mitigation efforts prove to be insufficient \citep{Keith2001,LentonVaughan2009,RoyalSociety2009,Kosugi2010}. This type of (solar) geoengineering is based on placing sun shields or reflecting particles in space between the Earth and the Sun \citep{Early1989,Angel2006,Fuglesangetal2021,Mautner1989,Pearson2006,McInnes2015}, or on the injection of aerosols into the stratosphere \citep{Crutzen2006,Raschetal2008}. General circulation models (GCMs) suggest that an insolation reduction of 3.5-5.0\% can largely undo the global temperature rise and intensified hydrological cycle associated with a quadrupled pre-industrial CO$_2$ concentration \citep{Luntetal2008,Kravitz2013,Bal2019,Irvine2016}. However, those GCMs also predict latitudinal variations in temperature response, and an extra reduction of precipitation in the tropics. Moreover, although clouds play a vital role in the Earth's radiation balance \citep{Ramanathan1989}, the impact of solar dimming on clouds is still poorly understood \citep{Schmidt2012,RussottoAckerman2018,Kravitz2021,VirginFletcher2022}. GCMs modeling the response to (extraterrestrial) dimming of sunlight are based on idealized scenarios \citep{KravitsGeoMIP2011}. They
inherently suffer from uncertainties \citep{Irvineetal2014}, mainly focus on the long-term impact, and highly parameterize short-term and small-scale processes such as cloud formation \citep{Irvineetal2014,Kravitz2021}. Twice a year on average, for a few hours the opportunity arises to take measurements of the Earth experiencing gradual insolation reductions from 0 to nearly 100\%, during the partial phase of a solar eclipse. Although the time scales involved with solar geoengineering will most likely not be equivalent to those of solar eclipses, these measurements can help to better understand (and test models that predict) the immediate cloud response to the deployment of sunlight intercepting material. 

\begin{figure*}[t!]%
\centering
\includegraphics[width=1.0\textwidth]{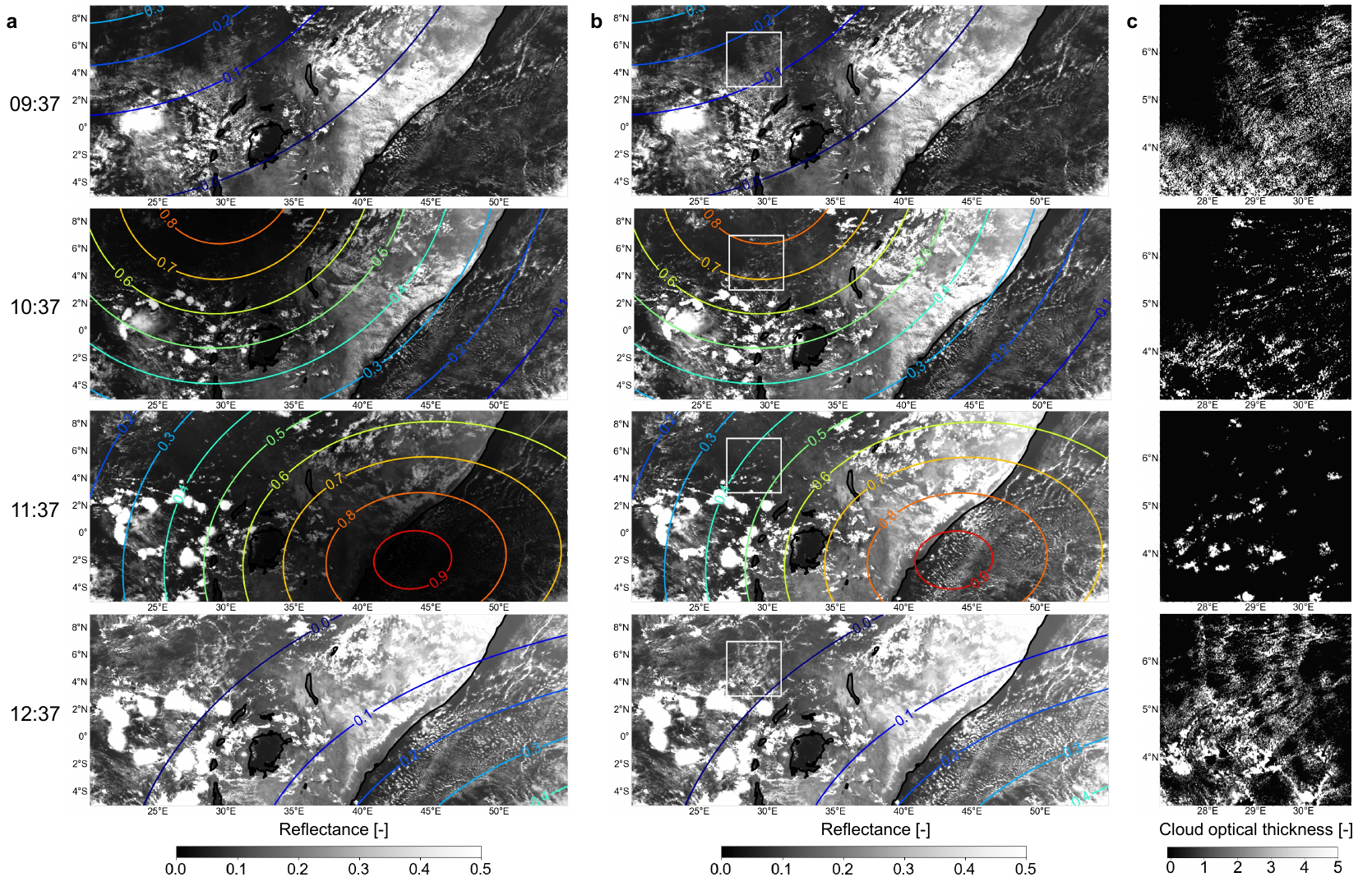}
\caption{SEVIRI images of the annular solar eclipse on 3 October 2005. (a) The original and (b) the eclipse corrected TOA VIS reflectance over East Africa (in the West) and Indian Ocean (in the East), at 09:37, 10:37, 11:37 and 12:37 UTC (from top to bottom). The colored contour lines indicate the solar obscuration fraction and the white squares over land mark the study area. (c) The corresponding eclipse corrected cloud optical thickness zoomed-in on the study area.}\label{fig1}
\end{figure*}

Ground-based meteorological observations during solar eclipses have primarily focused on fast drops in air temperature, winds and turbulence, and on changing (photo)chemistry \citep{Aplin2016,Harrison2016}. Weather publications contain anecdotal descriptions of dissipating low-level cumulus clouds right before totality, while mid- and high-level clouds survived  \citep{Anderson1999,Hanna2000}. The timing of the eclipse-induced effect on cumulus clouds is difficult to quantify with ground-based measurements due to the chaotic nature of cumulus cloud evolution and the missing observation of the non-eclipse state, and atmospheric model studies of eclipses did not yet analyze the cumulus cloud sensitivity \citep{Founda2007,Montornes2016,Clark2016,Buban2019}. Additionally, the precise locations of the affected clouds can be hard to predict and are sometimes inaccessible for ground-based observers. Earth observation satellites in geostationary orbit can continuously monitor clouds in large geographical areas \citep{Rossow1989,Stengel2014,Benas2017} and hints of dissipating cumulus clouds have been observed by comparing satellite images before and after a total solar eclipse, without estimating the non-eclipse state \citep{Anderson1999,Gerth2018}. However, during solar eclipses, satellite images show spatio-temporally varying darkening \citep{Gerth2018,Murillo2018} and satellite retrievals of cloud cover and cloud optical thickness (COT) are biased. This bias is caused by not taking into account the insolation reduction in the calculation of the top-of-atmosphere (TOA) reflectance from which cloud properties are derived. Hence, it has remained unknown how fast clouds are modified by various solar obscuration fractions.

Here, we present geostationary satellite measurements of clouds  
during three solar eclipses between 2005 and 2016 that have been corrected for the insolation reduction. Our corrected measurements reveal that shallow cumulus clouds start to dissipate at a solar obscuration of $\sim$15\% over cooling land surfaces, which would have been hidden in the partial lunar shadow without insolation reduction correction. Using large-eddy simulations we explain the timing of the cloud dissipation and demonstrate that the rising air parcels in the atmospheric boundary layer are already affected by the solar eclipse at even smaller obscurations. We calculate that neglecting the dissipating cloud behaviour in the simulations would result in an overestimation of 20 W m$^{-2}$ of the eclipse-related reduction of net incoming shortwave radiation at the top-of-atmosphere. Finally, we discuss that the high cloud sensitivity to rather small insolation reductions should spur cloud model simulations of the short-term impact of sunlight-intercepting geoengineering concepts.

\section*{Results}

\subsection*{Satellite observations}

Figure \ref{fig1} shows the uncorrected and corrected TOA visible (VIS) reflectance over East Africa and part of the Indian Ocean, at four subsequent hours during the annular solar eclipse on 3 October 2005, obtained by the Spinning Enhanced Visible and InfraRed Imager (SEVIRI) instrument (see Methods). In the first hour, cumulus clouds were present over land. They are better visible in the corrected high resolution images of the COT zoomed in on the specific area over land at 3$^\circ$-7$^\circ$ latitude and 27$^\circ$-31$^\circ$ longitude (Fig. \ref{fig1}c), which we call the study area. The derived COT in the cloudy pixels was $\lesssim$ 5 and SEVIRI estimated a cloud top altitude of 
$\sim$2 km (see Supplementary Figure \ref{figsup0}), indicating that they were shallow and located in the planetary boundary layer. Similar large-scale daytime shallow cumulus cloud fields over land can be 
found in central Africa and the Amazonian rainforest throughout the year, and in northeast America and Siberia during the boreal summer 
\citep{Rabin1996,Heiblum2014}. When the obscuration increased in the second hour, the cloud cover in the study area diminished, although remnants of the cloud pattern of the previous hour are still recognizable. In the third hour, the insolation increased again, but the shallow cumulus clouds stayed away while the clouds that survived had grown in size and COT. It was only during the final stage of the eclipse that shallow cumulus clouds returned throughout the study area. Over ocean, the cumulus clouds did not disappear. In Supplementary Figures \ref{figsup1} and \ref{figsup2}, we provide two more examples of vanishing shallow cumulus clouds over land during solar eclipses.

\begin{figure*}[t]
    \centering
        \includegraphics[width=0.9\textwidth]{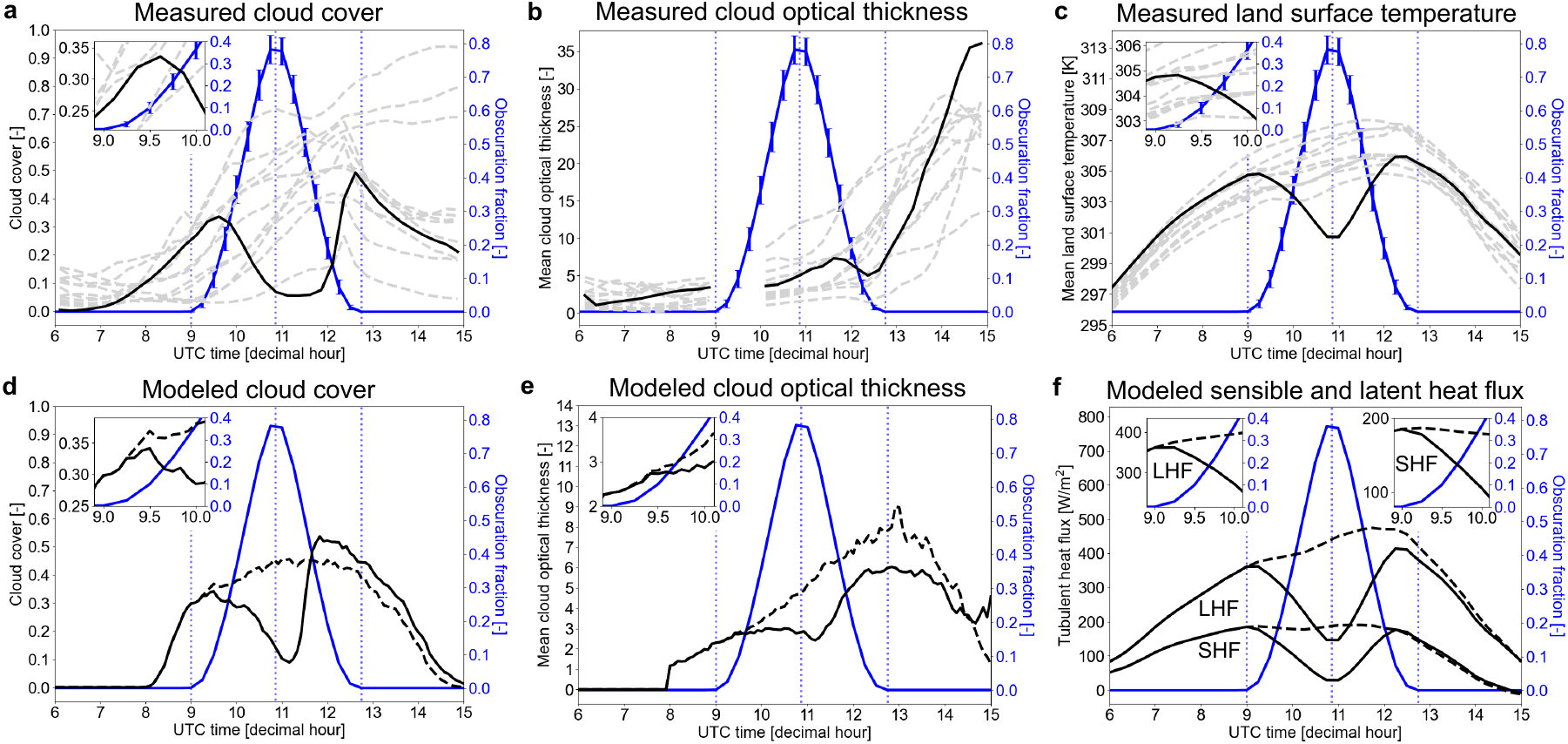} %
        \caption{Time series of cloud and land surface parameters in the study area (i.e., the white square in Fig. \ref{fig1}). (a) The retrieved cloud cover by SEVIRI. (b) The cloud optical thickness (COT) retrieved by SEVIRI. (c) The spatial mean land surface temperature (LST) retrieved by SEVIRI. (d) The modeled cloud cover with DALES. (e) The COT simulated with DALES. (f) The sensible and latent heat fluxes modeled with DALES. The black solid lines are for the day of the solar eclipse (3 October 2005). The grey dashed lines in (a), (b) and (c) are for the comparable days but without solar eclipse. The black dashed lines in (d), (e) and (f) are the results of the modeled reference case. The blue solid lines illustrate the spatial average obscuration fraction, with error bars in (a), (b) and (c) representing the standard deviation of the spatial variation. The blue vertical dotted lines indicate the start, maximum, and end of the eclipse (from left to right). The missing observations between 9 and 10 UTC in (b) are caused by COT retrieval errors at that scattering geometry due to the cloud bow. The insets show the time series zoomed in on 09:00 to 10:00 UTC. Local noon occurred at 10:04 UTC.} 
        \label{fig:timeseries}
\end{figure*}

The disappearance of shallow cumulus clouds only occurred on the day of the solar eclipse. Figure \ref{fig:timeseries}a shows the time series of the cloud cover in the study area, compared to that during 11 comparable days which we selected based on a similar type of cloud pattern before the eclipse started (see Methods). Also depicted in Fig. \ref{fig:timeseries} is the obscuration fraction on the eclipse day. The time series shows that the increasing cloud cover in the morning already halted at low obscuration fractions ($\sim$0.15), happening at around 09:30 UTC which was 30 minutes after the start of the eclipse. Secondly, there was a $\sim$50 minutes time lag with respect to the instant of maximum obscuration at 10:52 UTC before the clouds started to return.
During the increase in cloud cover between 12:00 and 12:30 UTC, the mean COT in the cloudy pixels decreased (see Fig. \ref{fig:timeseries}b), which can be attributed to the contribution of the newly formed shallow clouds. This decrease is absent in the time series of the comparable days.

Because shallow cumulus clouds in the boundary layer are generated by rising thermals originating from air close to the surface \citep{Siebesma1998}, we collected land surface temperature (LST) measurements from space by SEVIRI, derived from infrared radiation emitted by the shallow land surface layer (see Methods). Figure \ref{fig:timeseries}c shows the spatially averaged LST in the study area on the eclipse day and the comparable days. The maximum LST on the comparable days was two hours delayed with respect to local noon at 10:04 UTC, which could possibly be explained by the smaller heat flux into the ground due to the warmed subsurface layer in the afternoon \citep{Stull1988}. On the eclipse day, the LST drops instantly with the obscuration fraction, due to the direct response of the shallow land surface layer temperature to net radiation forcing \citep{Stull1988}. We estimate a maximum LST drop of 5.8 K induced by the eclipse at 11.00 UTC (see Methods). Comparable fast drops in satellite LST measurements have been found in a study over Europe during the total solar eclipse of 20 March 2015 by \citet{Good2016} who showed dependencies of the drop magnitude on the eclipse duration and time of the day (earlier eclipses gave larger drops), vegetation, surface height and distance to the coast. We did not detect a time lag in the LST minimum with respect to maximum solar obscuration: a time lag of $\sim$1.5 minutes as reported in literature \citep{Good2016} is not resolved in our LST data at 15 minutes intervals. Hence, the measured time lag of $\sim$50 minutes before clouds return in Fig. \ref{fig:timeseries}a cannot be explained by a time lag in the LST. Over ocean, we found no sea surface temperature drop when the eclipse passed (see Supplementary Figure \ref{figlstsst}), due to the large heat capacity of water and the efficient heat transport from the sea surface to deeper water layers through turbulent mixing \citep{Stull1988}.

\subsection*{Large-eddy simulations}

In order to explain the land-cloud interaction in the study area, we simulated the evolution of shallow cumulus clouds during a solar eclipse with the Dutch Atmospheric Large-Eddy Simulation (DALES) model \citep{Heus2010}. Figures \ref{fig:timeseries}d and \ref{fig:timeseries}e show the time series of the simulated cloud cover and mean COT, respectively, as would be measured from space (see Methods), and Supplementary Figure \ref{fig:cotlessnapshots} contains snapshots of the spatially resolved cloud fields. We present the results for the solar eclipse case, using the measured LST as input, and a reference case without eclipse-induced LST drop (see Methods). Indeed, in the solar eclipse case, our simulations show a substantial decrease in cloud cover with respect to the reference case. The cloud cover already differed $\sim$15 to $\sim$20 minutes after the start of the eclipse when the obscuration was still smaller than $10\%$. As in the observations (Fig. \ref{fig:timeseries}a), there is a time lag in the instant of minimum cloud cover with respect to mid-eclipse, after which the cloud cover rapidly increases. We note that the simulated cloud cover right after the eclipse is even larger than in the reference case, while the simulated COT is lower. A larger cloud cover after the eclipse compared to the hypothetical non-eclipse scenario is difficult to prove with our observations, due to the large variability of the cloud cover on the comparable days in the afternoon (cf. Fig. \ref{fig:timeseries}a). During the rapid increase in cloud cover, the simulated COT also increases, which is not in agreement with the observations, but can be attributed to the absence of the deeper convective clouds in the south part of the study area due to the horizontally averaged input settings in our simulations (see Methods).

\begin{figure*}[t!]%
\centering
\includegraphics[width=1.0\textwidth]{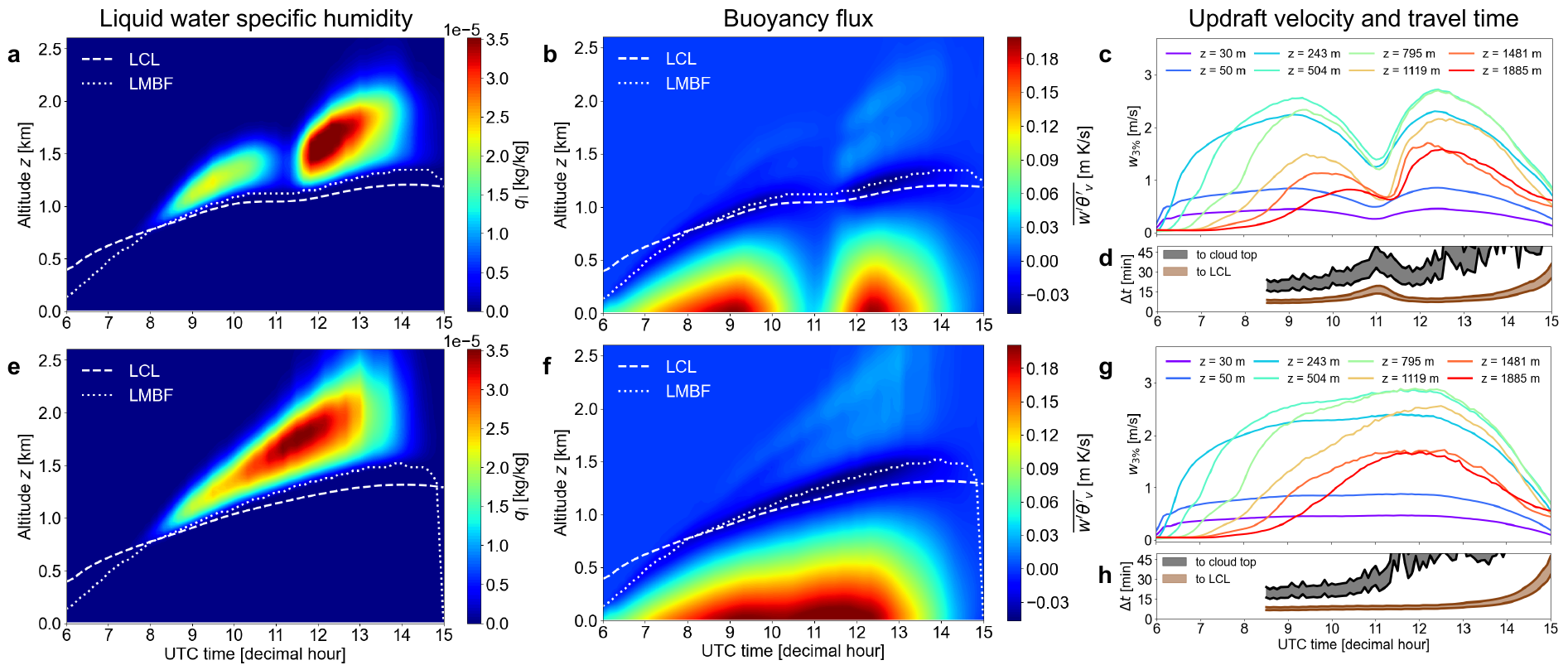}
\caption{Large-eddy simulation results from DALES for the solar eclipse and reference case. (a) Time-varying vertical profiles of the horizontal mean liquid water specific humidity $q_\mathrm{l}$ and (b) the buoyancy flux $\overline{w'\theta_\mathrm{v}'}$ for the solar eclipse case. (c) The vertical updraft velocity $w_\mathrm{3\%}$ of the 3-percentile fastest rising parcels at various altitudes and (d) the travel time $\Delta t$ of the 1- to 5-percentile fastest rising parcels from the surface to lifting condensation level (LCL) and to the cloud top (indicated with brown and grey color shades, respectively), for the solar eclipse case. Similar for Figs. (e), (f), (g) and (h), but then for the reference case. In Figs. (a), (b), (e) and (f), the LCL and level of minimum buoyancy flux (LMBF) are indicated with a dashed and dotted line, respectively.}\label{fig:contourplots}
\end{figure*}

The disappearing clouds during a solar eclipse can be explained by the drop in sensible (thermal) and latent (moisture) heat fluxes from the surface to the lowest atmosphere layer (Fig. \ref{fig:timeseries}f), as a result of the dropping LST (see Eqs. \ref{eq:shf} and \ref{eq:lhf}). Those heat fluxes drive the buoyancy flux (Figs. \ref{fig:contourplots}b and f) of relatively warm and moist air parcels from the surface, through the well-mixed boundary layer, up to the level of minimum buoyancy flux (LMBF) where the parcels are capped by a temperature inversion (see Supplementary Figure \ref{fig:profiles}). The rising parcels are cooled through adiabatic expansion which increases the parcel relatively humidity (RH) up to 100\% at the lifting condensation level (LCL) where shallow cumulus clouds are formed that can extend to higher altitudes (Figs. \ref{fig:contourplots}a and e). During a solar eclipse, this process is suppressed, which is clear from the overall drop in buoyancy flux. The drop in surface and buoyancy fluxes is consistent with the diminished boundary layer turbulence during solar eclipses found in other studies \citep{Eaton1997,Mauder2007,Buban2019}. It should be noted that the LMBF is still higher in our model than the LCL, but with smaller upward parcel velocities (Fig. \ref{fig:contourplots}c) which are controlled by the surface buoyancy flux, fewer parcels reach the LCL.

\begin{figure*}[t!]%
\centering
\includegraphics[width=0.6\textwidth]{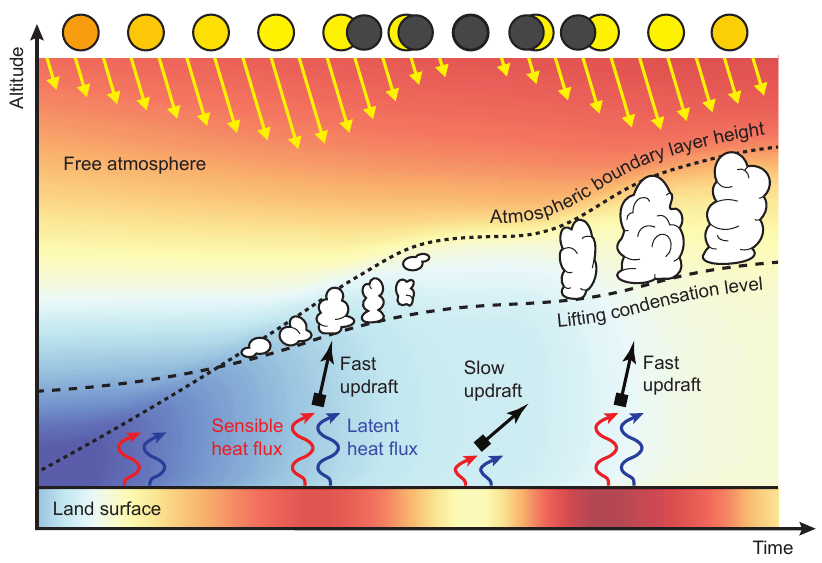}
\caption{Conceptual model of shallow cumulus cloud evolution during a solar eclipse. The time progresses in the horizontal direction to the right. Background color shading indicates the virtual potential temperature of the atmosphere and land surface in our simulation. The red and blue arrows are the sensible and latent heat fluxes, respectively, which depend strongly on the temperature difference between the surface and the atmosphere just above the surface. The yellow arrows represent the amount of incoming solar radiation, which is largest around noon but is reduced during a solar eclipse as illustrated by the lunar disk covering the solar disk. When the growing atmospheric boundary layer height (dotted line) intersects with the lifting condensation level (dashed line), clouds are formed, but they are diminished when the updrafts are slowed down during a solar eclipse, as indicated by the smaller inclination of the black arrow.}\label{fig:conceptualmodel}
\end{figure*}

The $\sim$15 to $\sim$20 minutes delay of the simulated vanishing cloud cover with respect to the dropping LST can be related to the $\sim$16 to $\sim$24 minutes travel time around 09:00 UTC of the fastest rising parcels from the lowest atmospheric layer to the cloud top as shown in Fig. \ref{fig:contourplots}d, after which the first individual clouds could fully disappear. We note that the vertical updraft velocity of the lowest layer responded within 5 minutes to differences in surface fluxes. The travel time depends on the cloud top height and vertical updraft velocities (see Figs. \ref{fig:contourplots}c and g and Methods). Thus, the cloud response was initiated when the parcels affected by the eclipse started rising, at even smaller obscuration fractions than at which this response could be observed. The $\sim$18 minutes time lag of the simulated cloud cover minimum with respect to mid-eclipse can be related to the relatively long travel time around 11:10 UTC of $\sim$13 to $\sim$19 minutes to LCL at which the newly formed clouds started influencing the cloud cover. Furthermore, we note that the LCL and cloud base remained slightly lower after the eclipse compared to the reference case, due to the continued colder air near the surface (see Supplementary Figure \ref{fig:botair} and Eq. \ref{eq:lawrence}). In Fig. \ref{fig:conceptualmodel} we provide a conceptual model in which we summarize the most important processes responsible for the shallow cumulus cloud behavior during a solar eclipse. 

\begin{figure}[t!]%
\centering
\includegraphics[width=1.0\columnwidth]{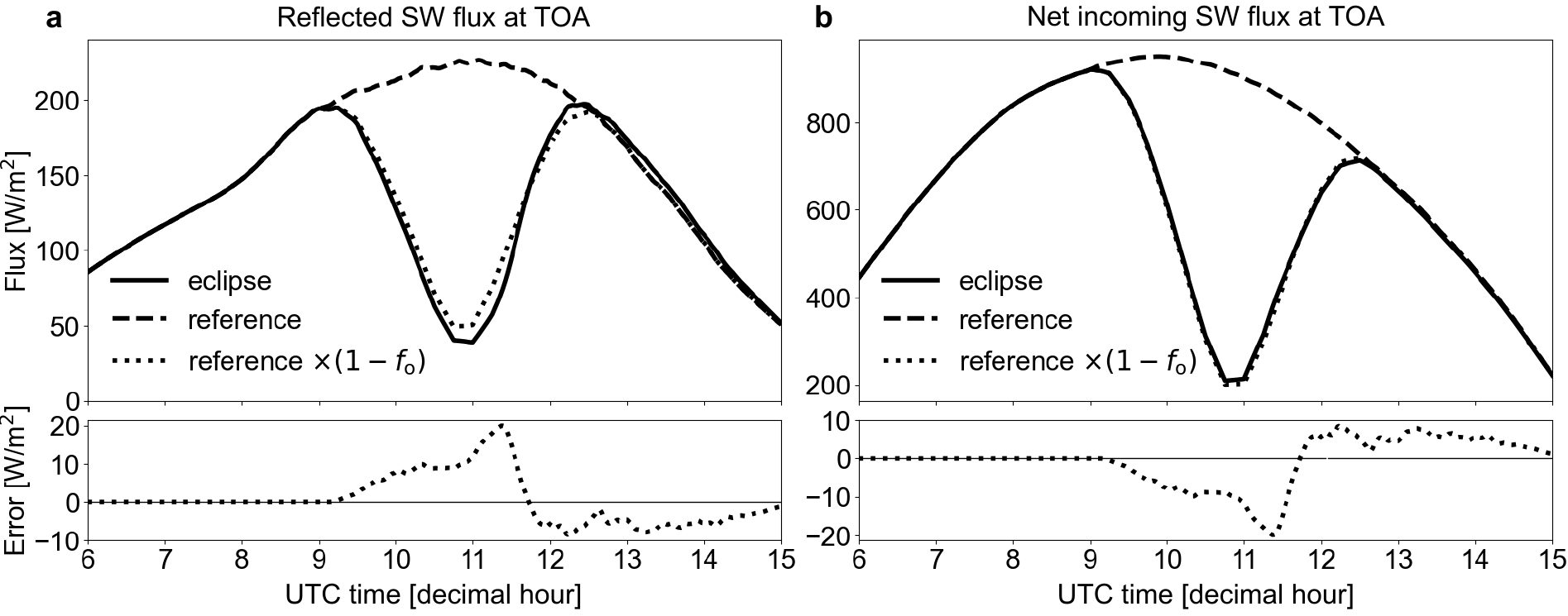}
\caption{Simulated radiative fluxes at the top-of-atmosphere (TOA). (a) The reflected and (b) the net incoming shortwave (SW) radiation (note the different scales of the vertical axes). The solid line is for the solar eclipse case, the dashed line is for the reference case, and the dotted line is for reference case multiplied by (1 - $f_\mathrm{o}$), with $f_\mathrm{o}$ the solar obscuration fraction. The bottom panels show the deviation of the latter from the solar eclipse case.}\label{fig:netswtoa}
\end{figure}

The disappearance of the shallow cumuli during the solar eclipse has a notable feedback on the solar radiative fluxes. This can be understood from Fig. \ref{fig:netswtoa}, which shows the simulated reflected and net incoming shortwave (SW) radiative flux at TOA, for both the reference and solar eclipse cases. As a result of the solar eclipse the net incoming SW flux started to decrease at 09:00 UTC. However, this change, which has solar dimming as its main cause, is also affected by the decrease of SW radiation reflected back into space due to the clearance of the sky. Indeed, the latter indirect effect causes an opposing increase in the net incoming SW flux at TOA. Neglecting the solar eclipse-induced cloud disappearance in our simulations (as illustrated by dotted line in Fig. \ref{fig:netswtoa}), resulted in an overestimation of 20 W m$^{-2}$ of the eclipse-related reduction of net incoming SW flux at TOA at 11:22 UTC. We note that this error would further increase with longer time lags of the cloud return with respect to mid-eclipse, such as found in the satellite observations (Fig \ref{fig:timeseries}a), because then more sunlight illuminates the cloud-free scenes.

\section*{Discussion}

The observed response of shallow cumulus clouds to a solar eclipse at already $\sim$15$\%$ obscuration, initiated at even smaller obscurations due to the parcel travel time, reveals the potential direct consequence of deploying sunlight intercepting material in the stratosphere or in space. We note that the duration of the cloud response is expected to depend on the speed and magnitude of the local obscuration variations, as the altered difference between the near-surface air and surface temperature, which causes the response, may possibly restore after a certain 
period. Diminished shallow cumulus clouds would partly oppose the objective of solar geoengineering which is to decrease the amount of net incoming solar radiation, and could prevent the growth into deeper convective and possibly precipitating clouds \citep{Grabowski2006}. While solar geoengineering proposals aim to reduce the solar radiation reaching the (lower) atmosphere and surface globally by only a few percent (depending on the required compensation), the use of non-uniform reductions to achieve this goal could increase the locally experienced variations in obscuration \citep{McInnes2015}. Space-based examples are the deployment of solar reflectors in Earth orbit \citep{Mautner1989,Pearson2006} or in orbit around the 1st Lagrange point \citep{McInnes2015}, offering daily and seasonally varying shading, respectively. Injected stratospheric aerosols can also exhibit spatio-temporally varying patterns due to seasonally changing global stratospheric circulation, depending on the injection location \citep{Tilmes2017} and strategy employed (whether constant or step-wise) \citep{Visioni2019,Lee2021}. Consequently, aerosol optical depths of 0.4-0.6 could be attained locally \citep{Visioni2019,Tilmes2017,Lee2021}, causing up to $\sim$45\% of the local direct sunlight to be scattered or absorbed by the aerosols before it reaches the lower atmosphere. Our results should spur model simulations investigating the response of shallow cumulus clouds to those geoengineering concepts, particularly for scenes over land where the surface temperature can adjust quickly. Additionally, our measurements provide an opportunity to validate these models, enhancing their reliability in predicting cloud behavior under natural conditions and in a world influenced by solar geoengineering.

\section*{Methods}\label{sec11}

\bmhead{Cloud measurements} 

The primary data set used in this research consists of measurements from the Spinning Enhanced Visible and InfraRed Imager (SEVIRI) on board the Meteosat Second Generation (MSG) series of geostationary satellites operated by EUMETSAT. We used data from Meteosat-8 and Meteosat-10, for the 2005-2006 and 2016 cases, respectively. SEVIRI measures TOA radiances over the full Earth disk centered at 0$^\circ$ latitude and 0$^\circ$ longitude every 15 minutes in twelve channels across the visible and near-infrared part of the spectrum \citep{Schmetz2002}. Eleven channels have a narrow bandwidth at low spatial resolution (3 x 3 km$^2$ at sub-satellite point) and one channel has a broad bandwidth (0.6 - 0.9 $\mu$m) at high spatial resolution (1 x 1 km$^2$), the latter being referred to as the HRV channel. Shortwave channel reflectances were obtained from the observed radiances and the calculated solar irradiance, and were calibrated with MODIS following \citet{Meirink2013}. For the longwave channels, the operational calibration from EUMETSAT was used.

For the surface temperature analyses at low spatial resolution (see Surface temperature measurements), cloud masks were calculated using the EUMETSAT Nowcasting and Very Short Range Forecasting Satellite Application Facility (NWC SAF) v2021 cloud algorithm \citep{NWCSAF2021}. The NWC SAF software also provided the cloud top height at low spatial resolution.
For the cloud analyses at high spatial resolution, we used the low-resolution NWC SAF cloud mask as a basis and improved the spatial resolution of the cloud mask using the HRV TOA reflectances by comparing with a HRV TOA reflectance climatology, which was generated for each pixel with HRV TOA reflectance measurements in a 16-day period centered at the day of the eclipse, following the method of \citet{Bley2013}. The TOA reflectances of the 0.6 and 1.6 $\mu$m channels were paired to simultaneously retrieve the COT and effective droplet radius \citep{Nakajima1990, Benas2017}, which were downscaled to high spatial resolution using the HRV channel as described in more detail in \citep{Deneke2021, Deneke2010}. The COT of the pixels that were not masked as cloudy was set equal to zero. For the study area (3-7$^\circ$ N latitude, 27-31$^\circ$ E longitude on 3 October 2005) and between 06:00 and 09:00 UTC, pixels with a TOA reflectance value larger than 1.0, originating from the sunglint in the rivers, were removed from the cloud product.

\bmhead{Solar eclipse correction}

We corrected the TOA reflectance for its reduction during a solar eclipse, that was due to the ignorance of the reduced solar irradiance in its calculation, through a division by 1 - $f_\mathrm{o}$, with $f_\mathrm{o}$ the solar obscuration fraction. In previous work \citep{Treesetal2021}, we validated this approach with the TROPOMI satellite instrument, which allowed for accurate monitoring of aerosols in the partial lunar shadow up to $f_\mathrm{o} = 0.92$. Recently, \citet{Wenetal2022} applied a similar type of correction to images of the Earth Polychromatic Imaging Camera (EPIC) instrument on the Deep Space Climate Observatory (DSCOVR), to quantify the measured TOA reflectance error without eclipse correction of the sunlit side of the Earth during the annular solar eclipse on 21 June 2020. The value of $f_\mathrm{o}$ is different for every pixel as it depends on measurement time, surface height, latitude and longitude, which are provided with the SEVIRI data. The so-called Besselian elements describing the temporal variation of the Moon shadow's geometry were taken from \citet{Espenak2006}. The value of $f_\mathrm{o}$ also depends on wavelength through the wavelength dependent limb darkening of the solar disk \citep{Koepke2001}. For the corrections of the TOA reflectances in the HRV channel and 0.6 $\mu$m, 0.8 $\mu$m and 1.6 $\mu$m channels, we used the limb darkening coefficients of \citet{PierceSlaughterSW1977} and \citet{PierceSlaughterLW1977} at 0.7 $\mu$m and the central wavelengths 0.635 $\mu$m, 0.81 $\mu$m and 1.64 $\mu$m, respectively. We refer to Trees et al. \citep{Treesetal2021} for more details about the solar eclipse correction of the TOA reflectance. Applying the cloud algorithms to the corrected TOA reflectance yielded the corrected cloud mask and COT.

\bmhead{Surface temperature measurements}

For the measurements of land surface temperature (LST), we used the LST product of the Land Surface Analysis Satellite Application Facility (LSA SAF) \citep{LST} derived from the 10.8 $\mu$m and 12 $\mu$m SEVIRI channels \citep{Wan1996} with an uncertainty of 1 to 2 K \citep{Freitas2010} which is expected to be stable within the time scales of solar eclipses \citep{Good2016}. We computed the horizontal average LST in the study area in every 15 minutes time step, after replacing the cloudy pixels by interpolated nearest neighbour values using the corrected SEVIRI NWC SAF cloud mask (see Cloud algorithm and Solar eclipse correction). The maximum LST drop due to the eclipse was estimated with respect to the average LST of comparable days (see Selection of comparable days). For the measurements of the sea surface temperature (SST), we used the hourly SST product of the Ocean and Sea Ice Satellite Application Facility (OSI SAF) \citep{SST} derived from the 10.8 $\mu$m and 12 $\mu$m SEVIRI channels with an uncertainty well below 1 K \citep{SST2020}.

\bmhead{Selection of comparable days}

The selection of days without solar eclipse that are comparable to our study case on 3 October 2005 in East Africa was done by first selecting 100 days from September and October in 2004, 2005 and 2006 with the smallest differences with respect to our study case in the sums of the predicted sensible and latent heat fluxes by ERA5 \citep{Hersbach2020} (which does not take into account the solar eclipse effect) in the complete diurnal cycle. Subsequently, we refined the selection by removing the days for which the mean of the absolute differences in high spatial resolution cloud cover with respect to our study case between 06:00 and 09:00 UTC was larger than the threshold of 0.1. In this way, we obtained 11 days for which we find the cloud cover comparable in the morning before the lunar shadow reached the study area: 2004-09-10, 2004-10-24, 2005-09-03, 2005-09-19, 2005-09-25, 2005-09-29, 2005-10-04, 2006-09-01, 2006-09-04, 2006-09-08, 2006-10-06. We visually inspected the HRV TOA reflectance and COT for those comparable days and indeed found between 06:00 and 09:00 UTC similar types of shallow cumulus clouds throughout the scene, with low COTs ($\lesssim$~5) as shown in Fig. \ref{fig:timeseries}b.

\bmhead{Cloud simulations}
The cloud simulations were performed with the Dutch Atmospheric Large-Eddy Simulation (DALES) model \citep{Heus2010}. The setup was a horizontally cyclic domain of 50 by 50 km$^2$ with horizontal cell sizes of 100 by 100 m$^2$ and 209 vertical layers between 0 and 14 km altitude. The vertical extent of each layer was stretched by a factor 1.01 with respect to the layer just below, and it was 20 m in the lowest layer. The ground surface was assumed flat. The atmosphere variables in the domain were initialized with horizontally homogeneous vertical profiles of the liquid potential temperature, total water specific humidity, and horizontal wind speed and direction. Those profiles were the horizontal mean profiles at 02:00 UTC in the study area (3$^\circ$-7$^\circ$N latitude, 27$^\circ$-31$^\circ$E longitude on 3 October 2005), taken from ERA5 \citep{Hersbach2020}. The pressure profile was determined with the use of the thermodynamic profiles, the gas law and mean hydrostatic balance. After 02:00 UTC, the simulation freely propagated those variables, until 15:00 UTC when the simulations ended. The time and altitude dependent mean horizontal advective tendencies of heat and moisture, in addition to the geostrophic winds, were all diagnosed from ERA5 and prescribed in the DALES runs. That is, we neglect the horizontal in- and outflow of eclipse-induced atmospheric changes at the boundaries of the domain, e.g. due to short-term disturbances in the horizontal pressure gradient \citep{Gross1999}. At the top boundary during the simulation, we imposed large-scale subsidence, also taken from ERA5.

The surface fluxes of heat and moisture were computed according to the vertical difference of potential temperature $\theta$ and the water vapor specific humidity $q_\mathrm{v}$ between the ground surface and the lowest atmospheric model layer (indicated by the subscript 'bot'), respectively. With the appropriate conversion of potential temperature to temperature the sensible and latent fluxes can be expressed as 
\begin{linenomath}
\begin{align}
    F_\mathrm{SH} = \frac{\rho c_\mathrm{p}}{r_{\mathrm{a,T}}}  (T_\mathrm{sfc} - T_\mathrm{bot} - \frac{gz_{\mathrm{bot}}}{c_\mathrm{p}} ) \label{eq:shf}\\
    F_\mathrm{LH} =  \frac{ \rho L_\mathrm{v}}{r_{\mathrm{a,q}}} (q_\mathrm{sat}(T_\mathrm{sfc}) - q_\mathrm{v,bot})
    \label{eq:lhf}
\end{align}
\end{linenomath}
\noindent where $\rho$ is the air density in kg m$^{-3}$,  $c_\mathrm{p} = 1004$ J kg$^{-1}$ K$^{-1}$ the heat capacity of dry air, $g = 9.81$ m s$^{-2}$ the gravitational acceleration, 
$L_\mathrm{v}= 2.5 \cdot 10^6$ J kg$^{-1}$ the latent heat for vaporization of water, and $q_\mathrm{sat}$ the temperature dependent saturation specific humidity. The aerodynamic resistance coefficients for  heat and moisture, $r_{\mathrm{a,T}}$ and $r_{\mathrm{a,q}}$, respectively, depend on the atmospheric stability near the surface following Monin-Obukhov similarity theory \citep{Heus2010}. The dependencies of the $r_\mathrm{a}$ values on the surface roughness length and the actual soil water content, which for land conditions will typically be much lower than the saturated value $q_\mathrm{sat}$, were taken into account by multiplying them with constant correction factors. This calibration step was made to obtain sensible and heat fluxes that were consistent with the ECMWF model for the reference case. As a consequence, the surface heat fluxes were not parameterized as functions of net incoming radiation, to avoid unnecessary complexity and uncertainties. Instead, the surface temperature $T_\mathrm{sfc}$ was prescribed using the SEVIRI LST measurements in the study area (see Satellite measurements), and shifted to the ERA5 skin temperature at 02:00 UTC through a positive offset of 2.0274 K applied to the complete LST time series, for consistency with the initial atmospheric profiles. In the simulation without eclipse and until 09:00 UTC, $T_\mathrm{sfc}$ was identical to that in the simulation with eclipse, but after 09:00 UTC it was the average of the LST time series of the comparable days shifted to the ERA5 skin temperature at 02:00 UTC. 

The simulated cloud cover in the study area was computed as follows. First, we regridded the DALES output liquid water specific humidity $q_\mathrm{l}$ in kg kg$^{-1}$ to a grid with cell sizes of 1 x 1 km$^2$, for a fair comparison with the SEVIRI observations. Secondly, we computed the COT by evaluating the following integral from the surface ($0$) to TOA ($z_\mathrm{TOA}$) \citep{Stephens1978}:
\begin{linenomath}
\begin{align}
\mathrm{COT}(x,y) = \int_{0}^{z_\mathrm{TOA}} \frac{\frac{3}{2} \rho_\mathrm{air}(z) q_\mathrm{l}(x,y,z)}{
 \rho_\mathrm{liq} r_\mathrm{eff}} dz
\end{align}
\end{linenomath}
\noindent where $\rho_\mathrm{air}$ is the air density in kg m$^{-3}$, $\rho_\mathrm{liq} = 1000$ kg m$^{-3}$ is the density of liquid water, and $r_\mathrm{eff}$ is the droplet effective radius which we assumed to be constant and equal to 10$^{-5}$ m. Thirdly, the columns ($x,y$) with a COT larger than 1 were flagged as a cloudy column. The cloud cover was computed as the number of cloudy columns divided by the total number of columns in the domain. For the mean COT time series, we computed the horizontal average COT considering the cloudy columns only.

The lifting condensation level (LCL) in m was computed from the DALES output of air temperature $T_\mathrm{bot}$ in K and relative humidity RH$_\mathrm{bot}$ in $\%$ of the bottom atmospheric layer (at 10 meter altitude), using the suggested formula by \citet{Lawrence2005}:
\begin{linenomath}
\begin{equation}
    z_\mathrm{LCL} = \left(20 + \frac{T_\mathrm{bot}-273.15}{5}\right)\cdot(100-\text{RH}_\mathrm{bot})
    \label{eq:lawrence}
\end{equation}
\end{linenomath}
\noindent The level of minimum buoyancy flux (LMBF) was computed as the altitude at which the buoyancy flux $\overline{w'\theta_v'}$ was minimum, where 
$\theta_\mathrm{v}$ is the virtual potential temperature which depends on specific humidity $q_\mathrm{v}$ in kg kg$^{-1}$, temperature $T$ in K, pressure $p$ in Pa, reference pressure $p_0 = 10^5$ Pa, $c_\mathrm{p}$, the gas constant for dry air $R_\mathrm{d} = 287.0$ J kg$^{-1}$ K$^{-1}$, and the gas constant for water vapor 
$R_\mathrm{v}$ = 461.5 J kg$^{-1}$ K$^{-1}$ \citep{CloudsandClimateSiebesma}:
\begin{linenomath}
\begin{equation}
    \theta_\mathrm{v} = T \left(\frac{p_0}
    {p}\right)^{\frac{R_\mathrm{d}}{c_\mathrm{p}}} \left(1 + \left(\frac{R_v}{R_d}-1\right) \cdot q_\mathrm{v} \right) 
\end{equation}
\end{linenomath}
and $\overline{w'\theta_\mathrm{v}'}$ was defined as:
\begin{linenomath}
\begin{equation}
\overline{w'\theta_\mathrm{v}'} = \frac{\sum^{N_x}_{i=1}\sum^{N_y}_{j=1}w_{ij}(\theta_{\mathrm{v,}ij}-\overline{\theta}_\mathrm{v})}{N_x N_y}
\end{equation}
\end{linenomath}
\noindent in which $N_x$ and $N_y$ are the number of grid cells in the horizontal $x$- and $y$ directions, and the overbar indicates a horizontal mean value. The mean updraft virtual potential temperature, $\overline{\theta}_\mathrm{v,up}(z)$, was computed as the mean of $\theta_\mathrm{v}(z)$ of the updrafts only (i.e., the grid points at a certain altitude $z$ for which $w_{ij} > 0$). The velocities of the fast rising parcels were defined by the $p$-percentile velocities $w_\mathrm{p\%}$ (with $p=1$, $3$ and $5$), which were the vertical velocities at a certain altitude $z$ for which a percentage $p$ of the $w$-distribution contained grid points with $w_{ij} > w_\mathrm{p\%}$ (see \citep{Siebesma2007}). The travel time $\Delta t$ of the fast rising parcels from the surface to a certain altitude $z_*$, using vertical velocity $w_\mathrm{p\%}$, was computed by numerically evaluating 
\begin{linenomath}
\begin{equation}
    \Delta t = \int_{0}^{z_\mathrm{*}} \frac{dz}{w_\mathrm{p\%}(z)}
\end{equation}
\end{linenomath}
\noindent with $z_\mathrm{*} = z_\mathrm{LCL}$ for the travel time to LCL and $z_\mathrm{*} = z_\mathrm{CT}$ for the travel time to the cloud top. The cloud top was defined as the highest altitude for which $q_\mathrm{l} > 0$.

DALES uses the {Rapid Radiative Transfer Model for Global climate model
applications (RRTMG) radiation scheme \citep{iacono2008radiative,blossey2013marine} to compute the shortwave (SW) and longwave (LW) radiative fluxes through the atmosphere, emitted and reflected by the surface, and emerging from TOA during the simulation. In  the solar eclipse case, the SW radiation incident at TOA was multiplied by ($1-f_\mathrm{o}$), where $f_\mathrm{o}$ was the horizontal mean obscuration fraction in the study area taken at 0.635 $\mu$m (see Solar eclipse correction). We used vertical profiles of the ozone mass mixing ratios in the study area from ERA5 \citep{Hersbach2020}. The surface albedo was the horizontal mean white-sky albedo in the study area measured by MODIS \citep{MODISalbedo}.

\bmhead{Supplementary information}

This article has accompanying supplementary videos of the corrected SEVIRI TOA VIS reflectance and cloud optical thickness during the solar eclipses of 3 October 2005, 29 March 2006, and 1 September 2016 (corresponding to Fig. \ref{fig1} and Supplementary Figures \ref{figsup1} and \ref{figsup2}, respectively) at 15 minutes temporal resolution.

\bmhead{Acknowledgments} The authors thank Richard Bintanja (KNMI) for the discussions during manuscript preparation and David Donovan (KNMI) for reading the manuscript before submission. VT acknowledges funding by the User Support Programme Space Research (GO), project number ALWGO.2018.016, of the Dutch Research Council (NWO). SdR acknowledges funding by the Refreeze the Arctic Foundation for research on geoengineering techniques at TU Delft.

\bmhead{Competing interests} The authors declare no competing interests.}

\bmhead{Author contributions} VT wrote the manuscript with contributions from SdR, JW and JFM. VT computed the solar obscuration fractions. SdR did the DALES computations. JW and JFM prepared the SEVIRI observations and JW converted the observations to high spatial resolution. VT, SdR, JW, JFM, PW, PS and APS reviewed the manuscript and were involved in the analysis and selection of the presented results.

\bmhead{Data availability}
The data required to replicate the timeseries presented in this article, as well as the input and boundary conditions for our model simulations, are available in the public repository Zenodo (\url{https://doi.org/10.5281/zenodo.10371414}).

\bmhead{Code availability}
The Dutch Atmospheric Large-Eddy Simulation (DALES) software used for this research is publicly available under the terms of the GNU GPL version 3 on \url{https://github.com/dalesteam/dales}.

\bibliography{sn-bibliography}


\begin{thebibliography}{75}
\ifx \bisbn   \undefined \def \bisbn  #1{ISBN #1}\fi
\ifx \binits  \undefined \def \binits#1{#1}\fi
\ifx \bauthor  \undefined \def \bauthor#1{#1}\fi
\ifx \batitle  \undefined \def \batitle#1{#1}\fi
\ifx \bjtitle  \undefined \def \bjtitle#1{#1}\fi
\ifx \bvolume  \undefined \def \bvolume#1{\textbf{#1}}\fi
\ifx \byear  \undefined \def \byear#1{#1}\fi
\ifx \bissue  \undefined \def \bissue#1{#1}\fi
\ifx \bfpage  \undefined \def \bfpage#1{#1}\fi
\ifx \blpage  \undefined \def \blpage #1{#1}\fi
\ifx \burl  \undefined \def \burl#1{\textsf{#1}}\fi
\ifx \doiurl  \undefined \def \doiurl#1{\url{https://doi.org/#1}}\fi
\ifx \betal  \undefined \def \betal{\textit{et al.}}\fi
\ifx \binstitute  \undefined \def \binstitute#1{#1}\fi
\ifx \binstitutionaled  \undefined \def \binstitutionaled#1{#1}\fi
\ifx \bctitle  \undefined \def \bctitle#1{#1}\fi
\ifx \beditor  \undefined \def \beditor#1{#1}\fi
\ifx \bpublisher  \undefined \def \bpublisher#1{#1}\fi
\ifx \bbtitle  \undefined \def \bbtitle#1{#1}\fi
\ifx \bedition  \undefined \def \bedition#1{#1}\fi
\ifx \bseriesno  \undefined \def \bseriesno#1{#1}\fi
\ifx \blocation  \undefined \def \blocation#1{#1}\fi
\ifx \bsertitle  \undefined \def \bsertitle#1{#1}\fi
\ifx \bsnm \undefined \def \bsnm#1{#1}\fi
\ifx \bsuffix \undefined \def \bsuffix#1{#1}\fi
\ifx \bparticle \undefined \def \bparticle#1{#1}\fi
\ifx \barticle \undefined \def \barticle#1{#1}\fi
\bibcommenthead
\ifx \bconfdate \undefined \def \bconfdate #1{#1}\fi
\ifx \botherref \undefined \def \botherref #1{#1}\fi
\ifx \url \undefined \def \url#1{\textsf{#1}}\fi
\ifx \bchapter \undefined \def \bchapter#1{#1}\fi
\ifx \bbook \undefined \def \bbook#1{#1}\fi
\ifx \bcomment \undefined \def \bcomment#1{#1}\fi
\ifx \oauthor \undefined \def \oauthor#1{#1}\fi
\ifx \citeauthoryear \undefined \def \citeauthoryear#1{#1}\fi
\ifx \endbibitem  \undefined \def \endbibitem {}\fi
\ifx \bconflocation  \undefined \def \bconflocation#1{#1}\fi
\ifx \arxivurl  \undefined \def \arxivurl#1{\textsf{#1}}\fi
\csname PreBibitemsHook\endcsname

\bibitem[\protect\citeauthoryear{{Keith}}{2001}]{Keith2001}
\begin{barticle}
\bauthor{\bsnm{{Keith}}, \binits{D.W.}}:
\batitle{{Geoengineering}}.
\bjtitle{Nature}
\bvolume{409}(\bissue{6818}),
\bfpage{420}
(\byear{2001})
\doiurl{10.1038/35053208}
\end{barticle}
\endbibitem

\bibitem[\protect\citeauthoryear{{Lenton} and
  {Vaughan}}{2009}]{LentonVaughan2009}
\begin{barticle}
\bauthor{\bsnm{{Lenton}}, \binits{T.M.}},
\bauthor{\bsnm{{Vaughan}}, \binits{N.E.}}:
\batitle{{The radiative forcing potential of different climate geoengineering
  options}}.
\bjtitle{Atmospheric Chemistry \& Physics}
\bvolume{9}(\bissue{15}),
\bfpage{5539}--\blpage{5561}
(\byear{2009})
\doiurl{10.5194/acp-9-5539-200910.5194/acpd-9-2559-2009}
\end{barticle}
\endbibitem

\bibitem[\protect\citeauthoryear{Shepherd}{2009}]{RoyalSociety2009}
\begin{botherref}
\oauthor{\bsnm{Shepherd}, \binits{J.G.}}:
Geoengineering the climate: science, governance and uncertainty.
Project report
(September 2009).
\url{https://eprints.soton.ac.uk/156647/}
\end{botherref}
\endbibitem

\bibitem[\protect\citeauthoryear{{Kosugi}}{2010}]{Kosugi2010}
\begin{barticle}
\bauthor{\bsnm{{Kosugi}}, \binits{T.}}:
\batitle{{Role of sunshades in space as a climate control option}}.
\bjtitle{Acta Astronautica}
\bvolume{67}(\bissue{1}),
\bfpage{241}--\blpage{253}
(\byear{2010})
\doiurl{10.1016/j.actaastro.2010.02.009}
\end{barticle}
\endbibitem

\bibitem[\protect\citeauthoryear{{Early}}{1989}]{Early1989}
\begin{barticle}
\bauthor{\bsnm{{Early}}, \binits{J.T.}}:
\batitle{{Space-based solar shield to offset greenhouse effect}}.
\bjtitle{Journal of the British Interplanetary Society}
\bvolume{42},
\bfpage{567}--\blpage{569}
(\byear{1989})
\end{barticle}
\endbibitem

\bibitem[\protect\citeauthoryear{{Angel}}{2006}]{Angel2006}
\begin{barticle}
\bauthor{\bsnm{{Angel}}, \binits{R.}}:
\batitle{{Feasibility of cooling the Earth with a cloud of small spacecraft
  near the inner Lagrange point (L1)}}.
\bjtitle{Proceedings of the National Academy of Science}
\bvolume{103}(\bissue{46}),
\bfpage{17184}--\blpage{17189}
(\byear{2006})
\doiurl{10.1073/pnas.0608163103}
\end{barticle}
\endbibitem

\bibitem[\protect\citeauthoryear{{Fuglesang} and {de Herreros
  Miciano}}{2021}]{Fuglesangetal2021}
\begin{barticle}
\bauthor{\bsnm{{Fuglesang}}, \binits{C.}},
\bauthor{\bsnm{{de Herreros Miciano}}, \binits{M.G.}}:
\batitle{{Realistic sunshade system at L$_{1}$ for global temperature
  control}}.
\bjtitle{Acta Astronautica}
\bvolume{186},
\bfpage{269}--\blpage{279}
(\byear{2021})
\doiurl{10.1016/j.actaastro.2021.04.035}
\end{barticle}
\endbibitem

\bibitem[\protect\citeauthoryear{{Mautner}}{1989}]{Mautner1989}
\begin{bchapter}
\bauthor{\bsnm{{Mautner}}, \binits{M.}}:
\bctitle{{A Space-Based Solar Screen Against Climatic Warming}}.
In: \bbtitle{Bulletin of the American Astronomical Society},
vol. \bseriesno{21},
p. \bfpage{993}
(\byear{1989})
\end{bchapter}
\endbibitem

\bibitem[\protect\citeauthoryear{{Pearson} et~al.}{2006}]{Pearson2006}
\begin{barticle}
\bauthor{\bsnm{{Pearson}}, \binits{J.}},
\bauthor{\bsnm{{Oldson}}, \binits{J.}},
\bauthor{\bsnm{{Levin}}, \binits{E.}}:
\batitle{{Earth rings for planetary environment control}}.
\bjtitle{Acta Astronautica}
\bvolume{58}(\bissue{1}),
\bfpage{44}--\blpage{57}
(\byear{2006})
\doiurl{10.1016/j.actaastro.2005.03.071}
\end{barticle}
\endbibitem

\bibitem[\protect\citeauthoryear{{S{\'a}nchez} and
  {McInnes}}{2015}]{McInnes2015}
\begin{barticle}
\bauthor{\bsnm{{S{\'a}nchez}}, \binits{J.-P.}},
\bauthor{\bsnm{{McInnes}}, \binits{C.R.}}:
\batitle{{Optimal Sunshade Configurations for Space-Based Geoengineering near
  the Sun-Earth L1 Point}}.
\bjtitle{PLoS ONE}
\bvolume{10}(\bissue{8}),
\bfpage{0136648}
(\byear{2015})
\doiurl{10.1371/journal.pone.0136648}
\end{barticle}
\endbibitem

\bibitem[\protect\citeauthoryear{{Crutzen}}{2006}]{Crutzen2006}
\begin{barticle}
\bauthor{\bsnm{{Crutzen}}, \binits{P.J.}}:
\batitle{{Albedo Enhancement by Stratospheric Sulfur Injections: A Contribution
  to Resolve a Policy Dilemma?}}
\bjtitle{Climatic Change}
\bvolume{77}(\bissue{3-4}),
\bfpage{211}--\blpage{220}
(\byear{2006})
\doiurl{10.1007/s10584-006-9101-y}
\end{barticle}
\endbibitem

\bibitem[\protect\citeauthoryear{{Rasch} et~al.}{2008}]{Raschetal2008}
\begin{barticle}
\bauthor{\bsnm{{Rasch}}, \binits{P.J.}},
\bauthor{\bsnm{{Tilmes}}, \binits{S.}},
\bauthor{\bsnm{{Turco}}, \binits{R.P.}},
\bauthor{\bsnm{{Robock}}, \binits{A.}},
\bauthor{\bsnm{{Oman}}, \binits{L.}},
\bauthor{\bsnm{{Chen}}, \binits{C.-C.J.}},
\bauthor{\bsnm{{Stenchikov}}, \binits{G.L.}},
\bauthor{\bsnm{{Garcia}}, \binits{R.R.}}:
\batitle{{An overview of geoengineering of climate using stratospheric sulphate
  aerosols}}.
\bjtitle{Philosophical Transactions of the Royal Society of London Series A}
\bvolume{366}(\bissue{1882}),
\bfpage{4007}--\blpage{4037}
(\byear{2008})
\doiurl{10.1098/rsta.2008.0131}
\end{barticle}
\endbibitem

\bibitem[\protect\citeauthoryear{{Lunt} et~al.}{2008}]{Luntetal2008}
\begin{barticle}
\bauthor{\bsnm{{Lunt}}, \binits{D.J.}},
\bauthor{\bsnm{{Ridgwell}}, \binits{A.}},
\bauthor{\bsnm{{Valdes}}, \binits{P.J.}},
\bauthor{\bsnm{{Seale}}, \binits{A.}}:
\batitle{{``Sunshade World'': A fully coupled GCM evaluation of the climatic
  impacts of geoengineering}}.
\bjtitle{Geophysical Research Letters}
\bvolume{35}(\bissue{12}),
\bfpage{12710}
(\byear{2008})
\doiurl{10.1029/2008GL033674}
\end{barticle}
\endbibitem

\bibitem[\protect\citeauthoryear{{Kravitz} et~al.}{2013}]{Kravitz2013}
\begin{barticle}
\bauthor{\bsnm{{Kravitz}}, \binits{B.}},
\bauthor{\bsnm{{Caldeira}}, \binits{K.}},
\bauthor{\bsnm{{Boucher}}, \binits{O.}},
\bauthor{\bsnm{{Robock}}, \binits{A.}},
\bauthor{\bsnm{{Rasch}}, \binits{P.J.}},
\bauthor{\bsnm{{Alterskj{\ae}r}}, \binits{K.}},
\bauthor{\bsnm{{Karam}}, \binits{D.B.}},
\bauthor{\bsnm{{Cole}}, \binits{J.N.S.}},
\bauthor{\bsnm{{Curry}}, \binits{C.L.}},
\bauthor{\bsnm{{Haywood}}, \binits{J.M.}},
\bauthor{\bsnm{{Irvine}}, \binits{P.J.}},
\bauthor{\bsnm{{Ji}}, \binits{D.}},
\bauthor{\bsnm{{Jones}}, \binits{A.}},
\bauthor{\bsnm{{Kristj{\'a}nsson}}, \binits{J.E.}},
\bauthor{\bsnm{{Lunt}}, \binits{D.J.}},
\bauthor{\bsnm{{Moore}}, \binits{J.C.}},
\bauthor{\bsnm{{Niemeier}}, \binits{U.}},
\bauthor{\bsnm{{Schmidt}}, \binits{H.}},
\bauthor{\bsnm{{Schulz}}, \binits{M.}},
\bauthor{\bsnm{{Singh}}, \binits{B.}},
\bauthor{\bsnm{{Tilmes}}, \binits{S.}},
\bauthor{\bsnm{{Watanabe}}, \binits{S.}},
\bauthor{\bsnm{{Yang}}, \binits{S.}},
\bauthor{\bsnm{{Yoon}}, \binits{J.-H.}}:
\batitle{{Climate model response from the Geoengineering Model Intercomparison
  Project (GeoMIP)}}.
\bjtitle{Journal of Geophysical Research (Atmospheres)}
\bvolume{118}(\bissue{15}),
\bfpage{8320}--\blpage{8332}
(\byear{2013})
\doiurl{10.1002/jgrd.50646}
\end{barticle}
\endbibitem

\bibitem[\protect\citeauthoryear{{Bal} et~al.}{2019}]{Bal2019}
\begin{barticle}
\bauthor{\bsnm{{Bal}}, \binits{P.K.}},
\bauthor{\bsnm{{Pathak}}, \binits{R.}},
\bauthor{\bsnm{{Mishra}}, \binits{S.K.}},
\bauthor{\bsnm{{Sahany}}, \binits{S.}}:
\batitle{{Effects of global warming and solar geoengineering on precipitation
  seasonality}}.
\bjtitle{Environmental Research Letters}
\bvolume{14}(\bissue{3}),
\bfpage{034011}
(\byear{2019})
\doiurl{10.1088/1748-9326/aafc7d}
\end{barticle}
\endbibitem

\bibitem[\protect\citeauthoryear{Irvine et~al.}{2016}]{Irvine2016}
\begin{barticle}
\bauthor{\bsnm{Irvine}, \binits{P.J.}},
\bauthor{\bsnm{Kravitz}, \binits{B.}},
\bauthor{\bsnm{Lawrence}, \binits{M.G.}},
\bauthor{\bsnm{Muri}, \binits{H.}}:
\batitle{An overview of the {E}arth system science of solar geoengineering}.
\bjtitle{WIREs Climate Change}
\bvolume{7}(\bissue{6}),
\bfpage{815}--\blpage{833}
(\byear{2016})
\doiurl{10.1002/wcc.423}
\end{barticle}
\endbibitem

\bibitem[\protect\citeauthoryear{{Ramanathan} et~al.}{1989}]{Ramanathan1989}
\begin{barticle}
\bauthor{\bsnm{{Ramanathan}}, \binits{V.}},
\bauthor{\bsnm{{Cess}}, \binits{R.D.}},
\bauthor{\bsnm{{Harrison}}, \binits{E.F.}},
\bauthor{\bsnm{{Minnis}}, \binits{P.}},
\bauthor{\bsnm{{Barkstrom}}, \binits{B.R.}},
\bauthor{\bsnm{{Ahmad}}, \binits{E.}},
\bauthor{\bsnm{{Hartmann}}, \binits{D.}}:
\batitle{{Cloud-Radiative Forcing and Climate: Results from the Earth Radiation
  Budget Experiment}}.
\bjtitle{Science}
\bvolume{243}(\bissue{4887}),
\bfpage{57}--\blpage{63}
(\byear{1989})
\doiurl{10.1126/science.243.4887.57}
\end{barticle}
\endbibitem

\bibitem[\protect\citeauthoryear{{Schmidt} et~al.}{2012}]{Schmidt2012}
\begin{barticle}
\bauthor{\bsnm{{Schmidt}}, \binits{H.}},
\bauthor{\bsnm{{Alterskj{\ae}r}}, \binits{K.}},
\bauthor{\bsnm{{Karam}}, \binits{D.B.}},
\bauthor{\bsnm{{Boucher}}, \binits{O.}},
\bauthor{\bsnm{{Jones}}, \binits{A.}},
\bauthor{\bsnm{{Kristj{\'a}nsson}}, \binits{J.E.}},
\bauthor{\bsnm{{Niemeier}}, \binits{U.}},
\bauthor{\bsnm{{Schulz}}, \binits{M.}},
\bauthor{\bsnm{{Aaheim}}, \binits{A.}},
\bauthor{\bsnm{{Benduhn}}, \binits{F.}},
\bauthor{\bsnm{{Lawrence}}, \binits{M.}},
\bauthor{\bsnm{{Timmreck}}, \binits{C.}}:
\batitle{{Solar irradiance reduction to counteract radiative forcing from a
  quadrupling of CO$_{2}$: climate responses simulated by four earth system
  models}}.
\bjtitle{Earth System Dynamics}
\bvolume{3}(\bissue{1}),
\bfpage{63}--\blpage{78}
(\byear{2012})
\doiurl{10.5194/esd-3-63-2012}
\end{barticle}
\endbibitem

\bibitem[\protect\citeauthoryear{{Russotto} and
  {Ackerman}}{2018}]{RussottoAckerman2018}
\begin{barticle}
\bauthor{\bsnm{{Russotto}}, \binits{R.D.}},
\bauthor{\bsnm{{Ackerman}}, \binits{T.P.}}:
\batitle{{Changes in clouds and thermodynamics under solar geoengineering and
  implications for required solar reduction}}.
\bjtitle{Atmospheric Chemistry \& Physics}
\bvolume{18}(\bissue{16}),
\bfpage{11905}--\blpage{11925}
(\byear{2018})
\doiurl{10.5194/acp-18-11905-2018}
\end{barticle}
\endbibitem

\bibitem[\protect\citeauthoryear{{Kravitz} et~al.}{2021}]{Kravitz2021}
\begin{barticle}
\bauthor{\bsnm{{Kravitz}}, \binits{B.}},
\bauthor{\bsnm{{MacMartin}}, \binits{D.G.}},
\bauthor{\bsnm{{Visioni}}, \binits{D.}},
\bauthor{\bsnm{{Boucher}}, \binits{O.}},
\bauthor{\bsnm{{Cole}}, \binits{J.N.S.}},
\bauthor{\bsnm{{Haywood}}, \binits{J.}},
\bauthor{\bsnm{{Jones}}, \binits{A.}},
\bauthor{\bsnm{{Lurton}}, \binits{T.}},
\bauthor{\bsnm{{Nabat}}, \binits{P.}},
\bauthor{\bsnm{{Niemeier}}, \binits{U.}},
\bauthor{\bsnm{{Robock}}, \binits{A.}},
\bauthor{\bsnm{{S{\'e}f{\'e}rian}}, \binits{R.}},
\bauthor{\bsnm{{Tilmes}}, \binits{S.}}:
\batitle{{Comparing different generations of idealized solar geoengineering
  simulations in the Geoengineering Model Intercomparison Project (GeoMIP)}}.
\bjtitle{Atmospheric Chemistry \& Physics}
\bvolume{21}(\bissue{6}),
\bfpage{4231}--\blpage{4247}
(\byear{2021})
\doiurl{10.5194/acp-21-4231-2021}
\end{barticle}
\endbibitem

\bibitem[\protect\citeauthoryear{{Virgin} and
  {Fletcher}}{2022}]{VirginFletcher2022}
\begin{barticle}
\bauthor{\bsnm{{Virgin}}, \binits{J.G.}},
\bauthor{\bsnm{{Fletcher}}, \binits{C.G.}}:
\batitle{{On the Linearity of External Forcing Response in Solar Geoengineering
  Experiments}}.
\bjtitle{Geophysical Research Letters}
\bvolume{49}(\bissue{15}),
\bfpage{00200}
(\byear{2022})
\doiurl{10.1029/2022GL100200}
\end{barticle}
\endbibitem

\bibitem[\protect\citeauthoryear{{Kravitz} et~al.}{2011}]{KravitsGeoMIP2011}
\begin{barticle}
\bauthor{\bsnm{{Kravitz}}, \binits{B.}},
\bauthor{\bsnm{{Robock}}, \binits{A.}},
\bauthor{\bsnm{{Boucher}}, \binits{O.}},
\bauthor{\bsnm{{Schmidt}}, \binits{H.}},
\bauthor{\bsnm{{Taylor}}, \binits{K.E.}},
\bauthor{\bsnm{{Stenchikov}}, \binits{G.}},
\bauthor{\bsnm{{Schulz}}, \binits{M.}}:
\batitle{{The Geoengineering Model Intercomparison Project (GeoMIP)}}.
\bjtitle{Atmospheric Science Letters}
\bvolume{12}(\bissue{2}),
\bfpage{162}--\blpage{167}
(\byear{2011})
\doiurl{10.1002/asl.316}
\end{barticle}
\endbibitem

\bibitem[\protect\citeauthoryear{{Irvine} et~al.}{2014}]{Irvineetal2014}
\begin{barticle}
\bauthor{\bsnm{{Irvine}}, \binits{P.J.}},
\bauthor{\bsnm{{Boucher}}, \binits{O.}},
\bauthor{\bsnm{{Kravitz}}, \binits{B.}},
\bauthor{\bsnm{{Alterskj{\ae}r}}, \binits{K.}},
\bauthor{\bsnm{{Cole}}, \binits{J.N.S.}},
\bauthor{\bsnm{{Ji}}, \binits{D.}},
\bauthor{\bsnm{{Jones}}, \binits{A.}},
\bauthor{\bsnm{{Lunt}}, \binits{D.J.}},
\bauthor{\bsnm{{Moore}}, \binits{J.C.}},
\bauthor{\bsnm{{Muri}}, \binits{H.}},
\bauthor{\bsnm{{Niemeier}}, \binits{U.}},
\bauthor{\bsnm{{Robock}}, \binits{A.}},
\bauthor{\bsnm{{Singh}}, \binits{B.}},
\bauthor{\bsnm{{Tilmes}}, \binits{S.}},
\bauthor{\bsnm{{Watanabe}}, \binits{S.}},
\bauthor{\bsnm{{Yang}}, \binits{S.}},
\bauthor{\bsnm{{Yoon}}, \binits{J.-H.}}:
\batitle{{Key factors governing uncertainty in the response to sunshade
  geoengineering from a comparison of the GeoMIP ensemble and a perturbed
  parameter ensemble}}.
\bjtitle{Journal of Geophysical Research (Atmospheres)}
\bvolume{119}(\bissue{13}),
\bfpage{7946}--\blpage{7962}
(\byear{2014})
\doiurl{10.1002/2013JD020716}
\end{barticle}
\endbibitem

\bibitem[\protect\citeauthoryear{{Aplin} et~al.}{2016}]{Aplin2016}
\begin{barticle}
\bauthor{\bsnm{{Aplin}}, \binits{K.L.}},
\bauthor{\bsnm{{Scott}}, \binits{C.J.}},
\bauthor{\bsnm{{Gray}}, \binits{S.L.}}:
\batitle{{Atmospheric changes from solar eclipses}}.
\bjtitle{Philosophical Transactions of the Royal Society of London Series A}
\bvolume{374}(\bissue{2077}),
\bfpage{20150217}
(\byear{2016})
\doiurl{10.1098/rsta.2015.0217}
{\href{https://arxiv.org/abs/1603.02987}{{arXiv:1603.02987}}}
\end{barticle}
\endbibitem

\bibitem[\protect\citeauthoryear{{Harrison} and {Hanna}}{2016}]{Harrison2016}
\begin{barticle}
\bauthor{\bsnm{{Harrison}}, \binits{R.G.}},
\bauthor{\bsnm{{Hanna}}, \binits{E.}}:
\batitle{{The solar eclipse: a natural meteorological experiment}}.
\bjtitle{Philosophical Transactions of the Royal Society of London Series A}
\bvolume{374}(\bissue{2077}),
\bfpage{20150225}
(\byear{2016})
\doiurl{10.1098/rsta.2015.0225}
\end{barticle}
\endbibitem

\bibitem[\protect\citeauthoryear{{Anderson}}{1999}]{Anderson1999}
\begin{barticle}
\bauthor{\bsnm{{Anderson}}, \binits{J.}}:
\batitle{{Meteorological changes during a solar eclipse}}.
\bjtitle{Weather}
\bvolume{54}(\bissue{7}),
\bfpage{207}--\blpage{215}
(\byear{1999})
\doiurl{10.1002/j.1477-8696.1999.tb06465.x}
\end{barticle}
\endbibitem

\bibitem[\protect\citeauthoryear{{Hanna}}{2000}]{Hanna2000}
\begin{barticle}
\bauthor{\bsnm{{Hanna}}, \binits{E.}}:
\batitle{{Meteorological effects of the solar eclipse of 11 August 1999}}.
\bjtitle{Weather}
\bvolume{55}(\bissue{12}),
\bfpage{430}--\blpage{446}
(\byear{2000})
\doiurl{10.1002/j.1477-8696.2000.tb06481.x}
\end{barticle}
\endbibitem

\bibitem[\protect\citeauthoryear{{Founda} et~al.}{2007}]{Founda2007}
\begin{barticle}
\bauthor{\bsnm{{Founda}}, \binits{D.}},
\bauthor{\bsnm{{Melas}}, \binits{D.}},
\bauthor{\bsnm{{Lykoudis}}, \binits{S.}},
\bauthor{\bsnm{{Lisaridis}}, \binits{I.}},
\bauthor{\bsnm{{Gerasopoulos}}, \binits{E.}},
\bauthor{\bsnm{{Kouvarakis}}, \binits{G.}},
\bauthor{\bsnm{{Petrakis}}, \binits{M.}},
\bauthor{\bsnm{{Zerefos}}, \binits{C.}}:
\batitle{{The effect of the total solar eclipse of 29 March 2006 on
  meteorological variables in Greece}}.
\bjtitle{Atmospheric Chemistry \& Physics}
\bvolume{7}(\bissue{21}),
\bfpage{5543}--\blpage{5553}
(\byear{2007})
\doiurl{10.5194/acp-7-5543-2007}
\end{barticle}
\endbibitem

\bibitem[\protect\citeauthoryear{Montorn\`es et~al.}{2016}]{Montornes2016}
\begin{barticle}
\bauthor{\bsnm{Montorn\`es}, \binits{A.}},
\bauthor{\bsnm{Codina}, \binits{B.}},
\bauthor{\bsnm{Zack}, \binits{J.W.}},
\bauthor{\bsnm{Sola}, \binits{Y.}}:
\batitle{Implementation of bessel's method for solar eclipses prediction in the
  wrf-arw model}.
\bjtitle{Atmospheric Chemistry and Physics}
\bvolume{16}(\bissue{9}),
\bfpage{5949}--\blpage{5967}
(\byear{2016})
\doiurl{10.5194/acp-16-5949-2016}
\end{barticle}
\endbibitem

\bibitem[\protect\citeauthoryear{{Clark}}{2016}]{Clark2016}
\begin{barticle}
\bauthor{\bsnm{{Clark}}, \binits{P.A.}}:
\batitle{{Numerical simulations of the impact of the 20 March 2015 eclipse on
  UK weather}}.
\bjtitle{Philosophical Transactions of the Royal Society of London Series A}
\bvolume{374}(\bissue{2077}),
\bfpage{20150218}
(\byear{2016})
\doiurl{10.1098/rsta.2015.0218}
\end{barticle}
\endbibitem

\bibitem[\protect\citeauthoryear{{Buban} et~al.}{2019}]{Buban2019}
\begin{barticle}
\bauthor{\bsnm{{Buban}}, \binits{M.S.}},
\bauthor{\bsnm{{Lee}}, \binits{T.R.}},
\bauthor{\bsnm{{Dumas}}, \binits{E.J.}},
\bauthor{\bsnm{{Baker}}, \binits{C.B.}},
\bauthor{\bsnm{{Heuer}}, \binits{M.}}:
\batitle{{Observations and Numerical Simulation of the Effects of the 21 August
  2017 North American Total Solar Eclipse on Surface Conditions and Atmospheric
  Boundary-Layer Evolution}}.
\bjtitle{Boundary-Layer Meteorology}
\bvolume{171}(\bissue{2}),
\bfpage{257}--\blpage{270}
(\byear{2019})
\doiurl{10.1007/s10546-018-00421-4}
\end{barticle}
\endbibitem

\bibitem[\protect\citeauthoryear{{Rossow}}{1989}]{Rossow1989}
\begin{barticle}
\bauthor{\bsnm{{Rossow}}, \binits{W.B.}}:
\batitle{{Measuring Cloud Properties from Space: A Review.}}
\bjtitle{Journal of Climate}
\bvolume{2}(\bissue{3}),
\bfpage{201}--\blpage{213}
(\byear{1989})
\doiurl{10.1175/1520-0442(1989)002<0201:MCPFSA>2.0.CO;2}
\end{barticle}
\endbibitem

\bibitem[\protect\citeauthoryear{{Stengel} et~al.}{2014}]{Stengel2014}
\begin{barticle}
\bauthor{\bsnm{{Stengel}}, \binits{M.S.}},
\bauthor{\bsnm{{Kniffka}}, \binits{A.K.}},
\bauthor{\bsnm{{Meirink}}, \binits{J.F.M.}},
\bauthor{\bsnm{{Lockhoff}}, \binits{M.L.}},
\bauthor{\bsnm{{Tan}}, \binits{J.T.}},
\bauthor{\bsnm{{Hollmann}}, \binits{R.H.}}:
\batitle{{CLAAS: the CM SAF cloud property data set using SEVIRI}}.
\bjtitle{Atmospheric Chemistry \& Physics}
\bvolume{14}(\bissue{8}),
\bfpage{4297}--\blpage{4311}
(\byear{2014})
\doiurl{10.5194/acp-14-4297-2014}
\end{barticle}
\endbibitem

\bibitem[\protect\citeauthoryear{{Benas} et~al.}{2017}]{Benas2017}
\begin{barticle}
\bauthor{\bsnm{{Benas}}, \binits{N.}},
\bauthor{\bsnm{{Finkensieper}}, \binits{S.}},
\bauthor{\bsnm{{Stengel}}, \binits{M.}},
\bauthor{\bsnm{{van Zadelhoff}}, \binits{G.-J.}},
\bauthor{\bsnm{{Hanschmann}}, \binits{T.}},
\bauthor{\bsnm{{Hollmann}}, \binits{R.}},
\bauthor{\bsnm{{Fokke Meirink}}, \binits{J.}}:
\batitle{{The MSG-SEVIRI-based cloud property data record CLAAS-2}}.
\bjtitle{Earth System Science Data}
\bvolume{9}(\bissue{2}),
\bfpage{415}--\blpage{434}
(\byear{2017})
\doiurl{10.5194/essd-9-415-2017}
\end{barticle}
\endbibitem

\bibitem[\protect\citeauthoryear{{Gerth}}{2018}]{Gerth2018}
\begin{barticle}
\bauthor{\bsnm{{Gerth}}, \binits{J.J.}}:
\batitle{{Shining light on sky cover during a total solar eclipse}}.
\bjtitle{Journal of Applied Remote Sensing}
\bvolume{12},
\bfpage{020501}
(\byear{2018})
\doiurl{10.1117/1.JRS.12.020501}
\end{barticle}
\endbibitem

\bibitem[\protect\citeauthoryear{{Pe{\~n}aloza-Murillo} and
  {Pasachoff}}{2018}]{Murillo2018}
\begin{barticle}
\bauthor{\bsnm{{Pe{\~n}aloza-Murillo}}, \binits{M.A.}},
\bauthor{\bsnm{{Pasachoff}}, \binits{J.M.}}:
\batitle{{Cloudiness and Solar Radiation During the Longest Total Solar Eclipse
  of the 21st Century at Tianhuangping (Zhejiang), China}}.
\bjtitle{Journal of Geophysical Research (Atmospheres)}
\bvolume{123}(\bissue{23}),
\bfpage{13443}--\blpage{13461}
(\byear{2018})
\doiurl{10.1029/2018JD029253}
\end{barticle}
\endbibitem

\bibitem[\protect\citeauthoryear{{Rabin} and {Martin}}{1996}]{Rabin1996}
\begin{barticle}
\bauthor{\bsnm{{Rabin}}, \binits{R.M.}},
\bauthor{\bsnm{{Martin}}, \binits{D.W.}}:
\batitle{{Satellite observations of shallow cumulus coverage over the central
  United States: An exploration of land use impact on cloud cover}}.
\bjtitle{Journal of Geophysical Research}
\bvolume{101}(\bissue{D3}),
\bfpage{7149}--\blpage{7155}
(\byear{1996})
\doiurl{10.1029/95JD02891}
\end{barticle}
\endbibitem

\bibitem[\protect\citeauthoryear{{Heiblum} et~al.}{2014}]{Heiblum2014}
\begin{barticle}
\bauthor{\bsnm{{Heiblum}}, \binits{R.H.}},
\bauthor{\bsnm{{Koren}}, \binits{I.}},
\bauthor{\bsnm{{Feingold}}, \binits{G.}}:
\batitle{{On the link between Amazonian forest properties and shallow cumulus
  cloud fields}}.
\bjtitle{Atmospheric Chemistry \& Physics}
\bvolume{14}(\bissue{12}),
\bfpage{6063}--\blpage{6074}
(\byear{2014})
\doiurl{10.5194/acp-14-6063-2014}
\end{barticle}
\endbibitem

\bibitem[\protect\citeauthoryear{Siebesma}{1998}]{Siebesma1998}
\begin{bbook}
\bauthor{\bsnm{Siebesma}, \binits{A.P.}}:
In: \beditor{\bsnm{Plate}, \binits{E.J.}},
\beditor{\bsnm{Fedorovich}, \binits{E.E.}},
\beditor{\bsnm{Viegas}, \binits{D.X.}},
\beditor{\bsnm{Wyngaard}, \binits{J.C.}} (eds.)
\bbtitle{Shallow Cumulus Convection},
pp. \bfpage{441}--\blpage{486}.
\bpublisher{Springer},
\blocation{Dordrecht}
(\byear{1998}).
\doiurl{10.1007/978-94-011-5058-3_19} .
\burl{https://doi.org/10.1007/978-94-011-5058-3_19}
\end{bbook}
\endbibitem

\bibitem[\protect\citeauthoryear{{Stull}}{1988}]{Stull1988}
\begin{bbook}
\bauthor{\bsnm{{Stull}}, \binits{R.B.}}:
\bbtitle{{An Introduction to Boundary Layer Meteorology}}.
\bpublisher{Kluwer Academic Publishers},
\blocation{Dordrecht, Boston and London}
(\byear{1988}).
\doiurl{10.1007/978-94-009-3027-8}
\end{bbook}
\endbibitem

\bibitem[\protect\citeauthoryear{{Good}}{2016}]{Good2016}
\begin{barticle}
\bauthor{\bsnm{{Good}}, \binits{E.}}:
\batitle{{Satellite observations of surface temperature during the March 2015
  total solar eclipse}}.
\bjtitle{Philosophical Transactions of the Royal Society of London Series A}
\bvolume{374}(\bissue{2077}),
\bfpage{20150219}
(\byear{2016})
\doiurl{10.1098/rsta.2015.0219}
\end{barticle}
\endbibitem

\bibitem[\protect\citeauthoryear{Heus et~al.}{2010}]{Heus2010}
\begin{barticle}
\bauthor{\bsnm{Heus}, \binits{T.}},
\bauthor{\bsnm{Heerwaarden}, \binits{C.C.}},
\bauthor{\bsnm{Jonker}, \binits{H.J.J.}},
\bauthor{\bsnm{Pier~Siebesma}, \binits{A.}},
\bauthor{\bsnm{Axelsen}, \binits{S.}},
\bauthor{\bsnm{Dries}, \binits{K.}},
\bauthor{\bsnm{Geoffroy}, \binits{O.}},
\bauthor{\bsnm{Moene}, \binits{A.F.}},
\bauthor{\bsnm{Pino}, \binits{D.}},
\bauthor{\bsnm{Roode}, \binits{S.R.}},
\bauthor{\bsnm{Arellano}, \binits{J.}}:
\batitle{Formulation of the dutch atmospheric large-eddy simulation (dales) and
  overview of its applications}.
\bjtitle{Geoscientific Model Development}
\bvolume{3}(\bissue{2}),
\bfpage{415}--\blpage{444}
(\byear{2010})
\doiurl{10.5194/gmd-3-415-2010}
\end{barticle}
\endbibitem

\bibitem[\protect\citeauthoryear{{Eaton} et~al.}{1997}]{Eaton1997}
\begin{barticle}
\bauthor{\bsnm{{Eaton}}, \binits{F.D.}},
\bauthor{\bsnm{{Hines}}, \binits{J.R.}},
\bauthor{\bsnm{{Hatch}}, \binits{W.H.}},
\bauthor{\bsnm{{Cionco}}, \binits{R.M.}},
\bauthor{\bsnm{{Byers}}, \binits{J.}},
\bauthor{\bsnm{{Garvey}}, \binits{D.}},
\bauthor{\bsnm{{Miller}}, \binits{D.R.}}:
\batitle{{Solar Eclipse Effects Observed in the Planetary Boundary Layer Over a
  Desert}}.
\bjtitle{Boundary-Layer Meteorology}
\bvolume{83}(\bissue{2}),
\bfpage{331}--\blpage{346}
(\byear{1997})
\doiurl{10.1023/A:1000219210055}
\end{barticle}
\endbibitem

\bibitem[\protect\citeauthoryear{{Mauder} et~al.}{2007}]{Mauder2007}
\begin{barticle}
\bauthor{\bsnm{{Mauder}}, \binits{M.}},
\bauthor{\bsnm{{Desjardins}}, \binits{R.L.}},
\bauthor{\bsnm{{Oncley}}, \binits{S.P.}},
\bauthor{\bsnm{{MacPherson}}, \binits{I.}}:
\batitle{{Atmospheric Response to a Partial Solar Eclipse over a Cotton Field
  in Central California}}.
\bjtitle{Journal of Applied Meteorology and Climatology}
\bvolume{46}(\bissue{11}),
\bfpage{1792}--\blpage{1803}
(\byear{2007})
\doiurl{10.1175/2007JAMC1495.1}
\end{barticle}
\endbibitem

\bibitem[\protect\citeauthoryear{{Grabowski} et~al.}{2006}]{Grabowski2006}
\begin{barticle}
\bauthor{\bsnm{{Grabowski}}, \binits{W.W.}},
\bauthor{\bsnm{{Bechtold}}, \binits{P.}},
\bauthor{\bsnm{{Cheng}}, \binits{A.}},
\bauthor{\bsnm{{Forbes}}, \binits{R.}},
\bauthor{\bsnm{{Halliwell}}, \binits{C.}},
\bauthor{\bsnm{{Khairoutdinov}}, \binits{M.}},
\bauthor{\bsnm{{Lang}}, \binits{S.}},
\bauthor{\bsnm{{Nasuno}}, \binits{T.}},
\bauthor{\bsnm{{Petch}}, \binits{J.}},
\bauthor{\bsnm{{Tao}}, \binits{W.K.}},
\bauthor{\bsnm{{Wong}}, \binits{R.}},
\bauthor{\bsnm{{Wu}}, \binits{X.}},
\bauthor{\bsnm{{Xu}}, \binits{K.M.}}:
\batitle{{Daytime convective development over land: A model intercomparison
  based on LBA observations}}.
\bjtitle{Quarterly Journal of the Royal Meteorological Society}
\bvolume{132}(\bissue{615}),
\bfpage{317}--\blpage{344}
(\byear{2006})
\doiurl{10.1256/qj.04.147}
\end{barticle}
\endbibitem

\bibitem[\protect\citeauthoryear{{Tilmes} et~al.}{2017}]{Tilmes2017}
\begin{barticle}
\bauthor{\bsnm{{Tilmes}}, \binits{S.}},
\bauthor{\bsnm{{Richter}}, \binits{J.H.}},
\bauthor{\bsnm{{Mills}}, \binits{M.J.}},
\bauthor{\bsnm{{Kravitz}}, \binits{B.}},
\bauthor{\bsnm{{MacMartin}}, \binits{D.G.}},
\bauthor{\bsnm{{Vitt}}, \binits{F.}},
\bauthor{\bsnm{{Tribbia}}, \binits{J.J.}},
\bauthor{\bsnm{{Lamarque}}, \binits{J.-F.}}:
\batitle{{Sensitivity of Aerosol Distribution and Climate Response to
  Stratospheric SO$_{2}$ Injection Locations}}.
\bjtitle{Journal of Geophysical Research (Atmospheres)}
\bvolume{122}(\bissue{23}),
\bfpage{12591}--\blpage{12615}
(\byear{2017})
\doiurl{10.1002/2017JD026888}
\end{barticle}
\endbibitem

\bibitem[\protect\citeauthoryear{{Visioni} et~al.}{2019}]{Visioni2019}
\begin{barticle}
\bauthor{\bsnm{{Visioni}}, \binits{D.}},
\bauthor{\bsnm{{MacMartin}}, \binits{D.G.}},
\bauthor{\bsnm{{Kravitz}}, \binits{B.}},
\bauthor{\bsnm{{Tilmes}}, \binits{S.}},
\bauthor{\bsnm{{Mills}}, \binits{M.J.}},
\bauthor{\bsnm{{Richter}}, \binits{J.H.}},
\bauthor{\bsnm{{Boudreau}}, \binits{M.P.}}:
\batitle{{Seasonal Injection Strategies for Stratospheric Aerosol
  Geoengineering}}.
\bjtitle{Geophysical Research Letters}
\bvolume{46}(\bissue{13}),
\bfpage{7790}--\blpage{7799}
(\byear{2019})
\doiurl{10.1029/2019GL083680}
\end{barticle}
\endbibitem

\bibitem[\protect\citeauthoryear{{Lee} et~al.}{2021}]{Lee2021}
\begin{barticle}
\bauthor{\bsnm{{Lee}}, \binits{W.R.}},
\bauthor{\bsnm{{MacMartin}}, \binits{D.G.}},
\bauthor{\bsnm{{Visioni}}, \binits{D.}},
\bauthor{\bsnm{{Kravitz}}, \binits{B.}}:
\batitle{{High Latitude Stratospheric Aerosol Geoengineering Can Be More
  Effective if Injection Is Limited to Spring}}.
\bjtitle{Geophysical Research Letters}
\bvolume{48}(\bissue{9}),
\bfpage{92696}
(\byear{2021})
\doiurl{10.1029/2021GL092696}
\end{barticle}
\endbibitem

\bibitem[\protect\citeauthoryear{{Schmetz} et~al.}{2002}]{Schmetz2002}
\begin{barticle}
\bauthor{\bsnm{{Schmetz}}, \binits{J.}},
\bauthor{\bsnm{{Pili}}, \binits{P.}},
\bauthor{\bsnm{{Tjemkes}}, \binits{S.}},
\bauthor{\bsnm{{Just}}, \binits{D.}},
\bauthor{\bsnm{{Kerkmann}}, \binits{J.}},
\bauthor{\bsnm{{Rota}}, \binits{S.}},
\bauthor{\bsnm{{Ratier}}, \binits{A.}}:
\batitle{{An Introduction to Meteosat Second Generation (MSG).}}
\bjtitle{Bulletin of the American Meteorological Society}
\bvolume{83}(\bissue{7}),
\bfpage{977}--\blpage{992}
(\byear{2002})
\doiurl{10.1175/1520-0477(2002)083<0977:AITMSG>2.3.CO;2}
\end{barticle}
\endbibitem

\bibitem[\protect\citeauthoryear{{Meirink} et~al.}{2013}]{Meirink2013}
\begin{barticle}
\bauthor{\bsnm{{Meirink}}, \binits{J.F.}},
\bauthor{\bsnm{{Roebeling}}, \binits{R.A.}},
\bauthor{\bsnm{{Stammes}}, \binits{P.}}:
\batitle{{Inter-calibration of polar imager solar channels using SEVIRI}}.
\bjtitle{Atmospheric Measurement Techniques}
\bvolume{6}(\bissue{9}),
\bfpage{2495}--\blpage{2508}
(\byear{2013})
\doiurl{10.5194/amt-6-2495-2013}
\end{barticle}
\endbibitem

\bibitem[\protect\citeauthoryear{Kerdraon and Fontaine}{2021}]{NWCSAF2021}
\begin{botherref}
\oauthor{\bsnm{Kerdraon}, \binits{G.}},
\oauthor{\bsnm{Fontaine}, \binits{E.}}:
Algorithm theoretical basis document for the cloud product processors of the
  nwc/geo (geo-cma-v5.1 (nwc-009), geo-ct-v4.1 (nwc-016), geo-ctth-v4.1
  (nwc-017) and geo-cmic-v2.1 (nwc-021).
Technical Report NWC/CDOP3/GEO/MFL/SCI/ATBD/Cloud, Issue 1, Rev 0.1,
Météo-France / Centre d’études en Météorologie Satellitaire
(October 2021)
\end{botherref}
\endbibitem

\bibitem[\protect\citeauthoryear{Bley and Deneke}{2013}]{Bley2013}
\begin{barticle}
\bauthor{\bsnm{Bley}, \binits{S.}},
\bauthor{\bsnm{Deneke}, \binits{H.}}:
\batitle{A threshold-based cloud mask for the high-resolution visible channel
  of {M}eteosat {S}econd {G}eneration {SEVIRI}}.
\bjtitle{Atmospheric Measurement Techniques}
\bvolume{6},
\bfpage{2713}--\blpage{2723}
(\byear{2013})
\doiurl{10.5194/amt-6-2713-2013}
\end{barticle}
\endbibitem

\bibitem[\protect\citeauthoryear{{Nakajima} and {King}}{1990}]{Nakajima1990}
\begin{barticle}
\bauthor{\bsnm{{Nakajima}}, \binits{T.}},
\bauthor{\bsnm{{King}}, \binits{M.D.}}:
\batitle{{Determination of the optical thickness and effective particle radius
  of clouds from reflected solar radiation measurements. I - Theory}}.
\bjtitle{Journal of Atmospheric Sciences}
\bvolume{47},
\bfpage{1878}--\blpage{1893}
(\byear{1990})
\doiurl{10.1175/1520-0469(1990)047<1878:DOTOTA>2.0.CO;2}
\end{barticle}
\endbibitem

\bibitem[\protect\citeauthoryear{Deneke et~al.}{2021}]{Deneke2021}
\begin{barticle}
\bauthor{\bsnm{Deneke}, \binits{H.}},
\bauthor{\bsnm{Barrientos-Velasco}, \binits{C.}},
\bauthor{\bsnm{Bley}, \binits{S.}},
\bauthor{\bsnm{Hunerbein}, \binits{A.}},
\bauthor{\bsnm{Lenk}, \binits{S.}},
\bauthor{\bsnm{Macke}, \binits{A.}},
\bauthor{\bsnm{Meirink}, \binits{J.F.}},
\bauthor{\bsnm{Schroedter-Homscheidt}, \binits{M.}},
\bauthor{\bsnm{Senf}, \binits{F.}},
\bauthor{\bsnm{Wang}, \binits{P.}},
\bauthor{\bsnm{Werner}, \binits{F.}},
\bauthor{\bsnm{Witthuhn}, \binits{J.}}:
\batitle{Increasing the spatial resolution of cloud property retrievals from
  {M}eteosat {SEVIRI} by use of its high-resolution visible channel:
  Implementation and examples}.
\bjtitle{Atmospheric Measurement Techniques}
\bvolume{14},
\bfpage{5107}--\blpage{5126}
(\byear{2021})
\doiurl{10.5194/amt-14-5107-2021}
\end{barticle}
\endbibitem

\bibitem[\protect\citeauthoryear{Deneke and Roebeling}{2010}]{Deneke2010}
\begin{barticle}
\bauthor{\bsnm{Deneke}, \binits{H.M.}},
\bauthor{\bsnm{Roebeling}, \binits{R.A.}}:
\batitle{Downscaling of {M}eteosat {SEVIRI} 0.6 and 0.8 $\mu$m channel
  radiances utilizing the high-resolution visible channel}.
\bjtitle{Atmospheric Chemistry and Physics}
\bvolume{10},
\bfpage{9761}--\blpage{9772}
(\byear{2010})
\doiurl{10.5194/acp-10-9761-2010}
\end{barticle}
\endbibitem

\bibitem[\protect\citeauthoryear{{Trees} et~al.}{2021}]{Treesetal2021}
\begin{barticle}
\bauthor{\bsnm{{Trees}}, \binits{V.}},
\bauthor{\bsnm{{Wang}}, \binits{P.}},
\bauthor{\bsnm{{Stammes}}, \binits{P.}}:
\batitle{{Restoring the top-of-atmosphere reflectance during solar eclipses: a
  proof of concept with the UV absorbing aerosol index measured by TROPOMI}}.
\bjtitle{Atmospheric Chemistry \& Physics}
\bvolume{21}(\bissue{11}),
\bfpage{8593}--\blpage{8614}
(\byear{2021})
\doiurl{10.5194/acp-21-8593-2021}
\end{barticle}
\endbibitem

\bibitem[\protect\citeauthoryear{Wen et~al.}{2022}]{Wenetal2022}
\begin{botherref}
\oauthor{\bsnm{Wen}, \binits{G.}},
\oauthor{\bsnm{Marshak}, \binits{A.}},
\oauthor{\bsnm{Herman}, \binits{J.}},
\oauthor{\bsnm{Wu}, \binits{D.}}:
Reduction of spectral radiance reflectance during the annular solar eclipse of
  21 june 2020 observed by epic.
Frontiers in Remote Sensing
\textbf{3}
(2022)
\doiurl{10.3389/frsen.2022.777314}
\end{botherref}
\endbibitem

\bibitem[\protect\citeauthoryear{{Espenak} and {Meeus}}{2006}]{Espenak2006}
\begin{bbook}
\bauthor{\bsnm{{Espenak}}, \binits{F.}},
\bauthor{\bsnm{{Meeus}}, \binits{J.}}:
\bbtitle{{Five Millennium Canon of Solar Eclipses : -1999 to +3000 (2000 BCE to
  3000 CE)}}.
\bpublisher{NASA Technical Publication, NASA Center for AeroSpace Information,
  TP-2006-214141},
\blocation{Hannover, Maryland, USA}
(\byear{2006})
\end{bbook}
\endbibitem

\bibitem[\protect\citeauthoryear{{Koepke} et~al.}{2001}]{Koepke2001}
\begin{barticle}
\bauthor{\bsnm{{Koepke}}, \binits{P.}},
\bauthor{\bsnm{{Reuder}}, \binits{J.}},
\bauthor{\bsnm{{Schween}}, \binits{J.}}:
\batitle{{Spectral variation of the solar radiation during an eclipse}}.
\bjtitle{Meteorologische Zeitschrift}
\bvolume{10}(\bissue{3}),
\bfpage{179}--\blpage{186}
(\byear{2001})
\doiurl{10.1127/0941-2948/2001/0010-0179}
\end{barticle}
\endbibitem

\bibitem[\protect\citeauthoryear{{Pierce} and
  {Slaughter}}{1977}]{PierceSlaughterSW1977}
\begin{barticle}
\bauthor{\bsnm{{Pierce}}, \binits{A.K.}},
\bauthor{\bsnm{{Slaughter}}, \binits{C.D.}}:
\batitle{{Solar limb darkening. I: lambda lambda (3033 - 7297).}}
\bjtitle{Solar Physics}
\bvolume{51}(\bissue{1}),
\bfpage{25}--\blpage{41}
(\byear{1977})
\doiurl{10.1007/BF00240442}
\end{barticle}
\endbibitem

\bibitem[\protect\citeauthoryear{{Pierce} et~al.}{1977}]{PierceSlaughterLW1977}
\begin{barticle}
\bauthor{\bsnm{{Pierce}}, \binits{A.K.}},
\bauthor{\bsnm{{Slaughter}}, \binits{C.D.}},
\bauthor{\bsnm{{Weinberger}}, \binits{D.}}:
\batitle{{Solar limb darkening in the interval 7404 - 24018 {\r{A}}, II.}}
\bjtitle{Solar Physics}
\bvolume{52}(\bissue{1}),
\bfpage{179}--\blpage{189}
(\byear{1977})
\doiurl{10.1007/BF00935800}
\end{barticle}
\endbibitem

\bibitem[\protect\citeauthoryear{{Trigo} et~al.}{2017}]{LST}
\begin{botherref}
\oauthor{\bsnm{{Trigo}}, \binits{I.}},
\oauthor{\bsnm{{Freitas}}, \binits{S.C.}},
\oauthor{\bsnm{{Bioucas-Dias}}, \binits{J.}},
\oauthor{\bsnm{{Barroso}}, \binits{C.}},
\oauthor{\bsnm{{Monteiro}}, \binits{I.T.}},
\oauthor{\bsnm{{Viterbo}}, \binits{P.}},
\oauthor{\bsnm{{Martins}}, \binits{J.P.}}:
Algorithm theoretical basis document for land surface temperature (lst).
Technical Report SAF/LAND/IM/ATBD\_MLST/1.2, Issue 2,
Land surface analysis satellite application facility (LSA-SAF)
(2017)
\end{botherref}
\endbibitem

\bibitem[\protect\citeauthoryear{{Wan} and {Dozier}}{1996}]{Wan1996}
\begin{barticle}
\bauthor{\bsnm{{Wan}}, \binits{Z.}},
\bauthor{\bsnm{{Dozier}}, \binits{J.}}:
\batitle{{A generalized split-window algorithm for retrieving land-surface
  temperature from space}}.
\bjtitle{IEEE Transactions on Geoscience and Remote Sensing}
\bvolume{34}(\bissue{4}),
\bfpage{892}--\blpage{905}
(\byear{1996})
\doiurl{10.1109/36.508406}
\end{barticle}
\endbibitem

\bibitem[\protect\citeauthoryear{{Freitas} et~al.}{2010}]{Freitas2010}
\begin{barticle}
\bauthor{\bsnm{{Freitas}}, \binits{S.C.}},
\bauthor{\bsnm{{Trigo}}, \binits{I.F.}},
\bauthor{\bsnm{{Bioucas-Dias}}, \binits{J.M.}},
\bauthor{\bsnm{{Gottsche}}, \binits{F.-M.}}:
\batitle{{Quantifying the Uncertainty of Land Surface Temperature Retrievals
  From SEVIRI/Meteosat}}.
\bjtitle{IEEE Transactions on Geoscience and Remote Sensing}
\bvolume{48}(\bissue{1}),
\bfpage{523}--\blpage{534}
(\byear{2010})
\doiurl{10.1109/TGRS.2009.2027697}
\end{barticle}
\endbibitem

\bibitem[\protect\citeauthoryear{{Saux-Picart}}{2018}]{SST}
\begin{botherref}
\oauthor{\bsnm{{Saux-Picart}}, \binits{S.}}:
Algorithm theoretical basis document for msg/seviri sea surface temperature
  data record, version 1.3.
Technical Report SAF/OSI/CDOP2/MF/SCI/MA/256,
Ocean and Sea Ice Satellite Application Facility (OSI SAF)
(2018).
\doiurl{10.15770/EUM_SAF_OSI_0004}
\end{botherref}
\endbibitem

\bibitem[\protect\citeauthoryear{{Saux Picart} et~al.}{2020}]{SST2020}
\begin{barticle}
\bauthor{\bsnm{{Saux Picart}}, \binits{S.}},
\bauthor{\bsnm{{Marsouin}}, \binits{A.}},
\bauthor{\bsnm{{Legendre}}, \binits{G.}},
\bauthor{\bsnm{{Roquet}}, \binits{H.}},
\bauthor{\bsnm{{P{\'e}r{\'e}}}, \binits{S.}},
\bauthor{\bsnm{{Nano-Ascione}}, \binits{N.}},
\bauthor{\bsnm{{Gianelli}}, \binits{T.}}:
\batitle{{A Sea Surface Temperature data record (2004-2012) from Meteosat
  Second Generation satellites}}.
\bjtitle{Remote Sensing of Environment}
\bvolume{240},
\bfpage{111687}
(\byear{2020})
\doiurl{10.1016/j.rse.2020.111687}
\end{barticle}
\endbibitem

\bibitem[\protect\citeauthoryear{Hersbach et~al.}{2020}]{Hersbach2020}
\begin{barticle}
\bauthor{\bsnm{Hersbach}, \binits{H.}},
\bauthor{\bsnm{Bell}, \binits{B.}},
\bauthor{\bsnm{Berrisford}, \binits{P.}},
\bauthor{\bsnm{Hirahara}, \binits{S.}},
\bauthor{\bsnm{Horányi}, \binits{A.}},
\bauthor{\bsnm{Muñoz-Sabater}, \binits{J.}},
\bauthor{\bsnm{Nicolas}, \binits{J.}},
\bauthor{\bsnm{Peubey}, \binits{C.}},
\bauthor{\bsnm{Radu}, \binits{R.}},
\bauthor{\bsnm{Schepers}, \binits{D.}},
\bauthor{\bsnm{Simmons}, \binits{A.}},
\bauthor{\bsnm{Soci}, \binits{C.}},
\bauthor{\bsnm{Abdalla}, \binits{S.}},
\bauthor{\bsnm{Abellan}, \binits{X.}},
\bauthor{\bsnm{Balsamo}, \binits{G.}},
\bauthor{\bsnm{Bechtold}, \binits{P.}},
\bauthor{\bsnm{Biavati}, \binits{G.}},
\bauthor{\bsnm{Bidlot}, \binits{J.}},
\bauthor{\bsnm{Bonavita}, \binits{M.}},
\bauthor{\bsnm{De~Chiara}, \binits{G.}},
\bauthor{\bsnm{Dahlgren}, \binits{P.}},
\bauthor{\bsnm{Dee}, \binits{D.}},
\bauthor{\bsnm{Diamantakis}, \binits{M.}},
\bauthor{\bsnm{Dragani}, \binits{R.}},
\bauthor{\bsnm{Flemming}, \binits{J.}},
\bauthor{\bsnm{Forbes}, \binits{R.}},
\bauthor{\bsnm{Fuentes}, \binits{M.}},
\bauthor{\bsnm{Geer}, \binits{A.}},
\bauthor{\bsnm{Haimberger}, \binits{L.}},
\bauthor{\bsnm{Healy}, \binits{S.}},
\bauthor{\bsnm{Hogan}, \binits{R.J.}},
\bauthor{\bsnm{Hólm}, \binits{E.}},
\bauthor{\bsnm{Janisková}, \binits{M.}},
\bauthor{\bsnm{Keeley}, \binits{S.}},
\bauthor{\bsnm{Laloyaux}, \binits{P.}},
\bauthor{\bsnm{Lopez}, \binits{P.}},
\bauthor{\bsnm{Lupu}, \binits{C.}},
\bauthor{\bsnm{Radnoti}, \binits{G.}},
\bauthor{\bsnm{Rosnay}, \binits{P.}},
\bauthor{\bsnm{Rozum}, \binits{I.}},
\bauthor{\bsnm{Vamborg}, \binits{F.}},
\bauthor{\bsnm{Villaume}, \binits{S.}},
\bauthor{\bsnm{Thépaut}, \binits{J.-N.}}:
\batitle{The era5 global reanalysis}.
\bjtitle{Quarterly Journal of the Royal Meteorological Society}
\bvolume{146}(\bissue{730}),
\bfpage{1999}--\blpage{2049}
(\byear{2020})
\doiurl{10.1002/qj.3803}
\end{barticle}
\endbibitem

\bibitem[\protect\citeauthoryear{{Gross} and {Hense}}{1999}]{Gross1999}
\begin{barticle}
\bauthor{\bsnm{{Gross}}, \binits{P.}},
\bauthor{\bsnm{{Hense}}, \binits{A.}}:
\batitle{{Effects of a Total Solar Eclipse on the Mesoscale Atmospheric
  Circulation over Europe - A Model Experiment}}.
\bjtitle{Meteorology and Atmospheric Physics}
\bvolume{71},
\bfpage{229}--\blpage{242}
(\byear{1999})
\doiurl{10.1007/s007030050057}
\end{barticle}
\endbibitem

\bibitem[\protect\citeauthoryear{{Stephens}}{1978}]{Stephens1978}
\begin{barticle}
\bauthor{\bsnm{{Stephens}}, \binits{G.L.}}:
\batitle{{Radiation Profiles in Extended Water Clouds. II: Parameterization
  Schemes.}}
\bjtitle{Journal of Atmospheric Sciences}
\bvolume{35}(\bissue{11}),
\bfpage{2123}--\blpage{2132}
(\byear{1978})
\doiurl{10.1175/1520-0469(1978)035<2123:RPIEWC>2.0.CO;2}
\end{barticle}
\endbibitem

\bibitem[\protect\citeauthoryear{{Lawrence}}{2005}]{Lawrence2005}
\begin{barticle}
\bauthor{\bsnm{{Lawrence}}, \binits{M.G.}}:
\batitle{{The Relationship between Relative Humidity and the Dewpoint
  Temperature in Moist Air: A Simple Conversion and Applications.}}
\bjtitle{Bulletin of the American Meteorological Society}
\bvolume{86}(\bissue{2}),
\bfpage{225}--\blpage{233}
(\byear{2005})
\doiurl{10.1175/BAMS-86-2-225}
\end{barticle}
\endbibitem

\bibitem[\protect\citeauthoryear{}{2020}]{CloudsandClimateSiebesma}
\begin{botherref}
Clouds and Climate: Climate Science's Greatest Challenge.
Cambridge University Press
(2020).
\doiurl{10.1017/9781107447738}
\end{botherref}
\endbibitem

\bibitem[\protect\citeauthoryear{{Siebesma} et~al.}{2007}]{Siebesma2007}
\begin{barticle}
\bauthor{\bsnm{{Siebesma}}, \binits{A.P.}},
\bauthor{\bsnm{{Soares}}, \binits{P.M.M.}},
\bauthor{\bsnm{{Teixeira}}, \binits{J.}}:
\batitle{{A Combined Eddy-Diffusivity Mass-Flux Approach for the Convective
  Boundary Layer}}.
\bjtitle{Journal of Atmospheric Sciences}
\bvolume{64}(\bissue{4}),
\bfpage{1230}
(\byear{2007})
\doiurl{10.1175/JAS3888.1}
\end{barticle}
\endbibitem

\bibitem[\protect\citeauthoryear{{Iacono} et~al.}{2008}]{iacono2008radiative}
\begin{barticle}
\bauthor{\bsnm{{Iacono}}, \binits{M.J.}},
\bauthor{\bsnm{{Delamere}}, \binits{J.S.}},
\bauthor{\bsnm{{Mlawer}}, \binits{E.J.}},
\bauthor{\bsnm{{Shephard}}, \binits{M.W.}},
\bauthor{\bsnm{{Clough}}, \binits{S.A.}},
\bauthor{\bsnm{{Collins}}, \binits{W.D.}}:
\batitle{{Radiative forcing by long-lived greenhouse gases: Calculations with
  the AER radiative transfer models}}.
\bjtitle{Journal of Geophysical Research (Atmospheres)}
\bvolume{113}(\bissue{D13}),
\bfpage{13103}
(\byear{2008})
\doiurl{10.1029/2008JD009944}
\end{barticle}
\endbibitem

\bibitem[\protect\citeauthoryear{{Blossey} et~al.}{2013}]{blossey2013marine}
\begin{barticle}
\bauthor{\bsnm{{Blossey}}, \binits{P.N.}},
\bauthor{\bsnm{{Bretherton}}, \binits{C.S.}},
\bauthor{\bsnm{{Zhang}}, \binits{M.}},
\bauthor{\bsnm{{Cheng}}, \binits{A.}},
\bauthor{\bsnm{{Endo}}, \binits{S.}},
\bauthor{\bsnm{{Heus}}, \binits{T.}},
\bauthor{\bsnm{{Liu}}, \binits{Y.}},
\bauthor{\bsnm{{Lock}}, \binits{A.P.}},
\bauthor{\bsnm{{de Roode}}, \binits{S.R.}},
\bauthor{\bsnm{{Xu}}, \binits{K.-M.}}:
\batitle{{Marine low cloud sensitivity to an idealized climate change: The
  CGILS LES intercomparison}}.
\bjtitle{Journal of Advances in Modeling Earth Systems}
\bvolume{5}(\bissue{2}),
\bfpage{234}--\blpage{258}
(\byear{2013})
\doiurl{10.1002/jame.20025}
\end{barticle}
\endbibitem

\bibitem[\protect\citeauthoryear{Schaaf and Wang}{2015}]{MODISalbedo}
\begin{botherref}
\oauthor{\bsnm{Schaaf}, \binits{C.}},
\oauthor{\bsnm{Wang}, \binits{Z.}}:
MCD43C3 MODIS/Terra+Aqua BRDF/Albedo Albedo Daily L3 Global 0.05Deg CMG V006
  Data set.
NASA EOSDIS Land Processes DAAC, accessed 2023-06-24
  \url{https://doi.org/10.5067/MODIS/MCD43C3.006},
(2015)
\end{botherref}
\endbibitem

\end{thebibliography}

\newpage

\setcounter{figure}{0}

\makeatletter
\renewcommand{\fnum@figure}{Supplementary \figurename ~\thefigure}
\makeatother

\begin{figure*}[t!]%
\centering
\includegraphics[width=0.35\textwidth]{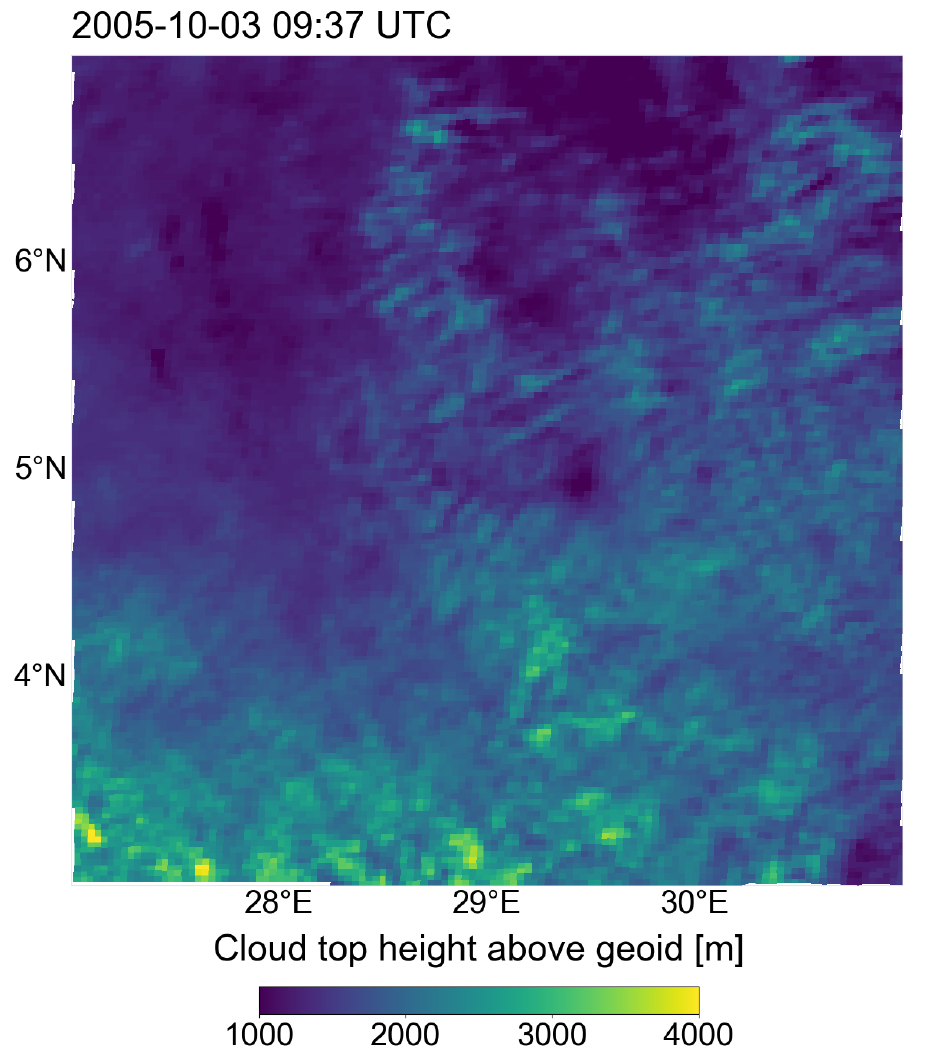}
\caption{Cloud top height above geoid in the study area on 3 October 2005 measured by SEVIRI. The surface height above geoid in this area varies between 300 and 1700 m.}\label{figsup0}
\end{figure*}

\begin{figure*}[t!]%
\centering
\includegraphics[width=1.0\textwidth]{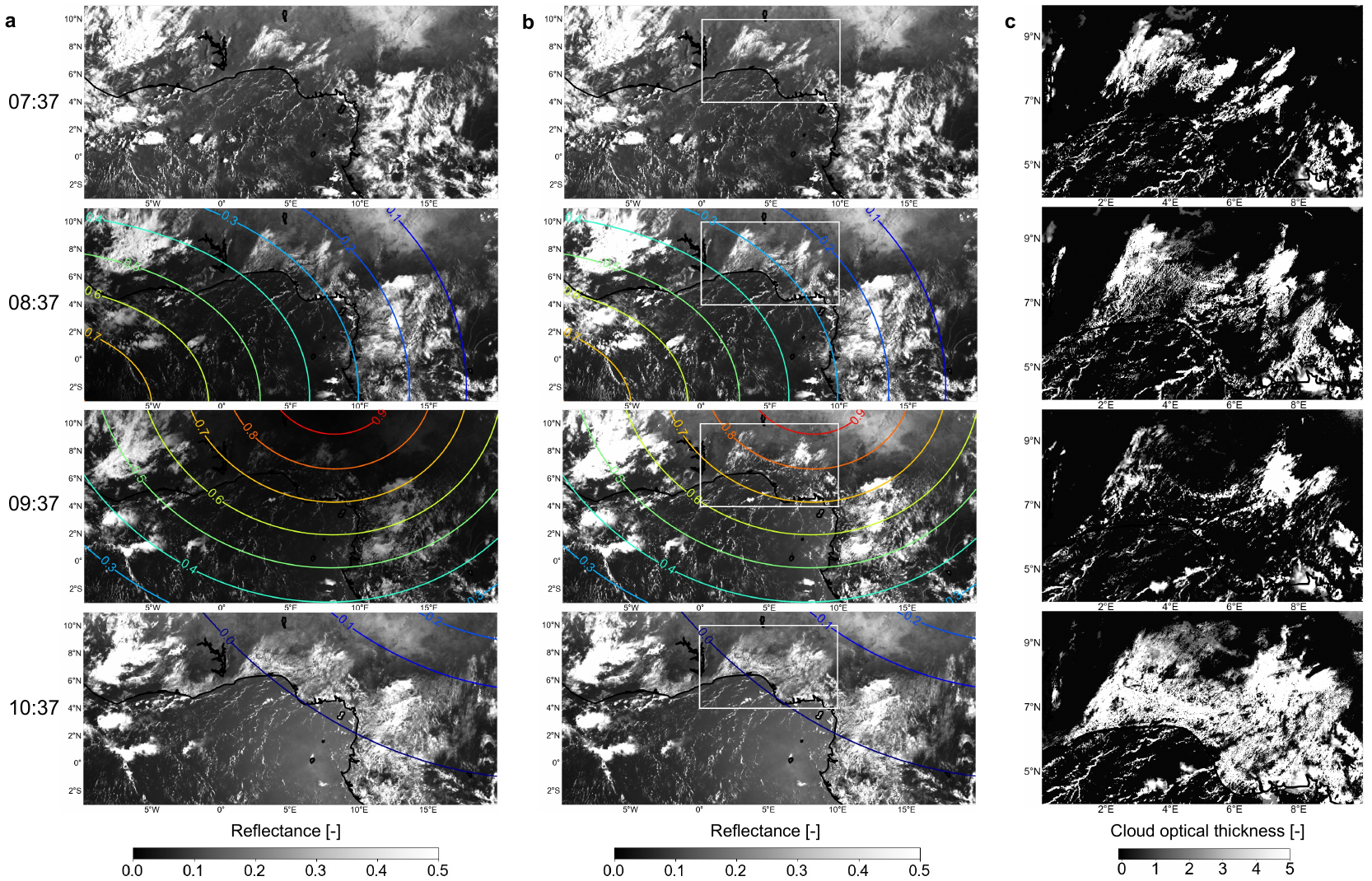}
\caption{Similar to Fig. \ref{fig1}, but for the total solar eclipse of 29 March 2006 passing over West Africa (in the North-East) and the Atlantic ocean (in the South-West) at 07:37, 08:37, 09:37 and 10:37 UTC (from top to bottom). The white squares in (b) indicate the zoom area for the cloud optical thickness in (c).}\label{figsup1}
\end{figure*}

\begin{figure*}[t!]%
\centering
\includegraphics[width=1.0\textwidth]{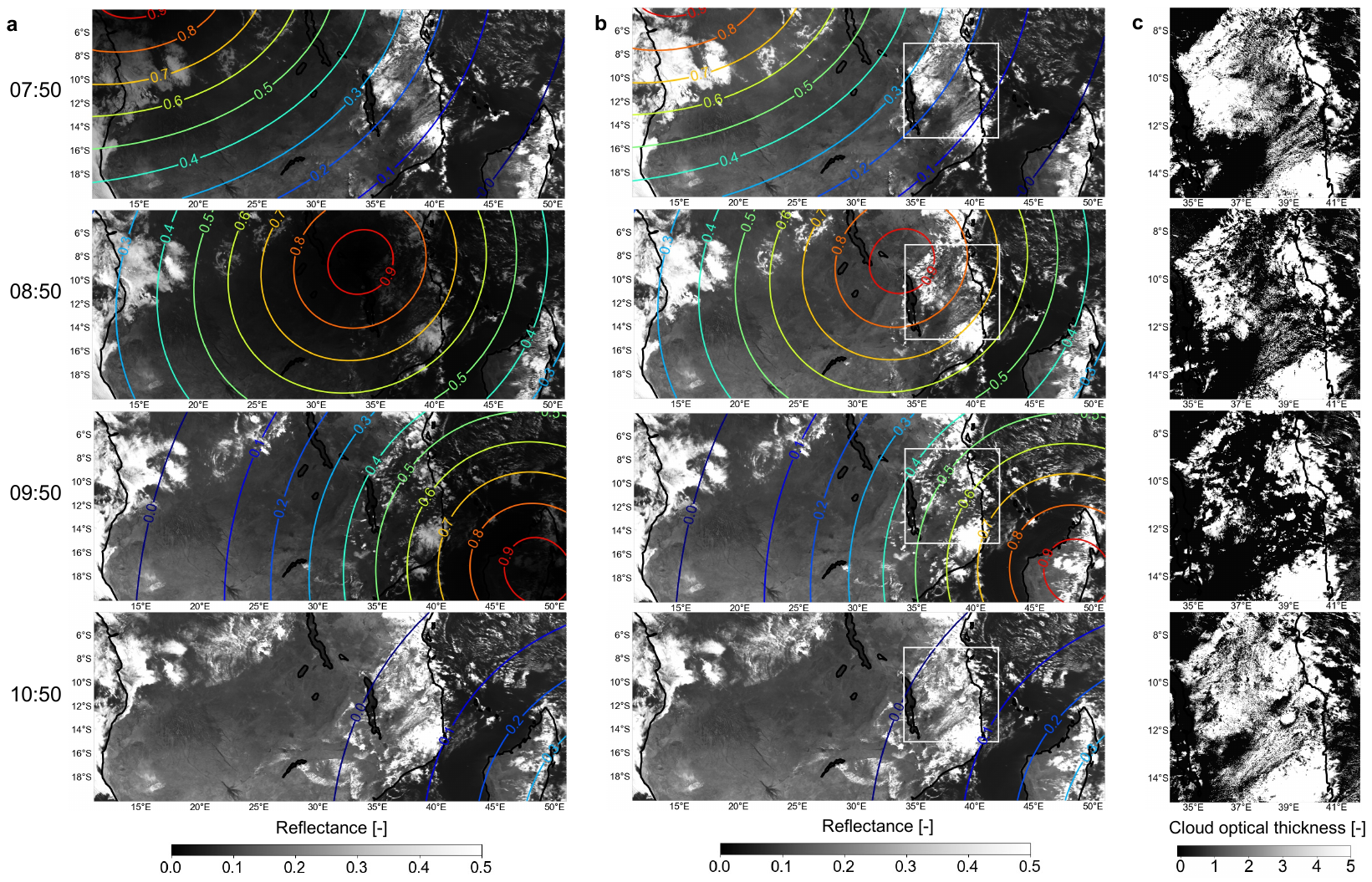}
\caption{Similar to Fig. \ref{fig1}, but for the annular solar eclipse of 1 September 2016 passing over Middle Africa and the Indian Ocean (in the East) at 07:50, 08:50, 09:50 and 10:50 UTC (from top to bottom). The white squares in (b) indicate the zoom area for the cloud optical thickness in (c).}\label{figsup2}
\end{figure*}

\begin{figure*}[t!]%
\centering
\includegraphics[width=1.0\textwidth]{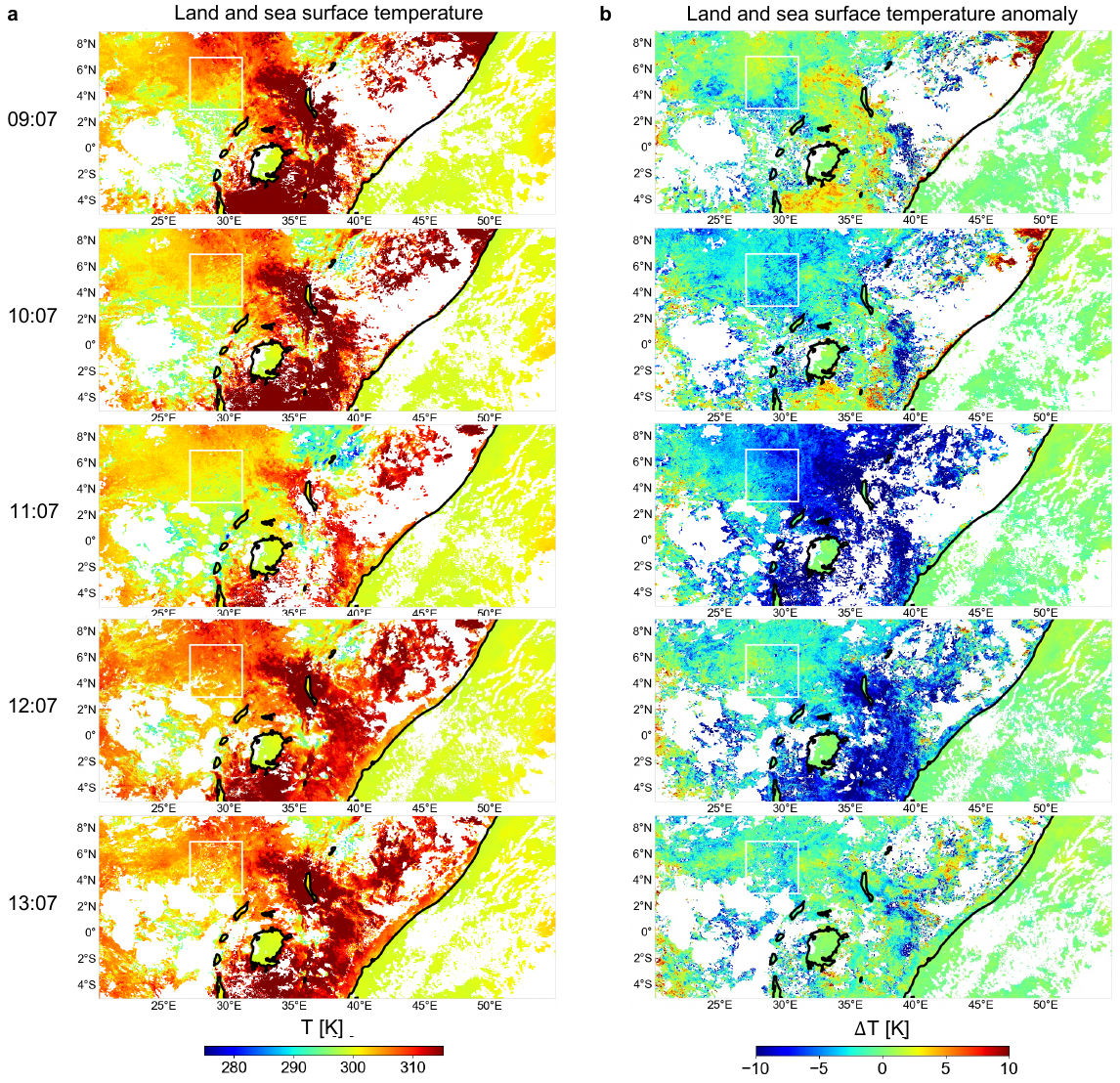}
\caption{SEVIRI images of the land and sea surface temperature and their anomaly over East Africa and part of the Indian Ocean during the annular solar eclipse on 3 October 2015. (a) The land surface temperature measurements at five subsequent hours (from top to bottom) and (b) the deviation from the mean of the comparable days. The white square indicates the study area. The negative deviation over land at 10:07, 11:07 and 12:07 UTC was caused by the solar eclipse.}\label{figlstsst}
\end{figure*}

\begin{figure*}[t!]%
\centering
\includegraphics[width=0.7\textwidth]{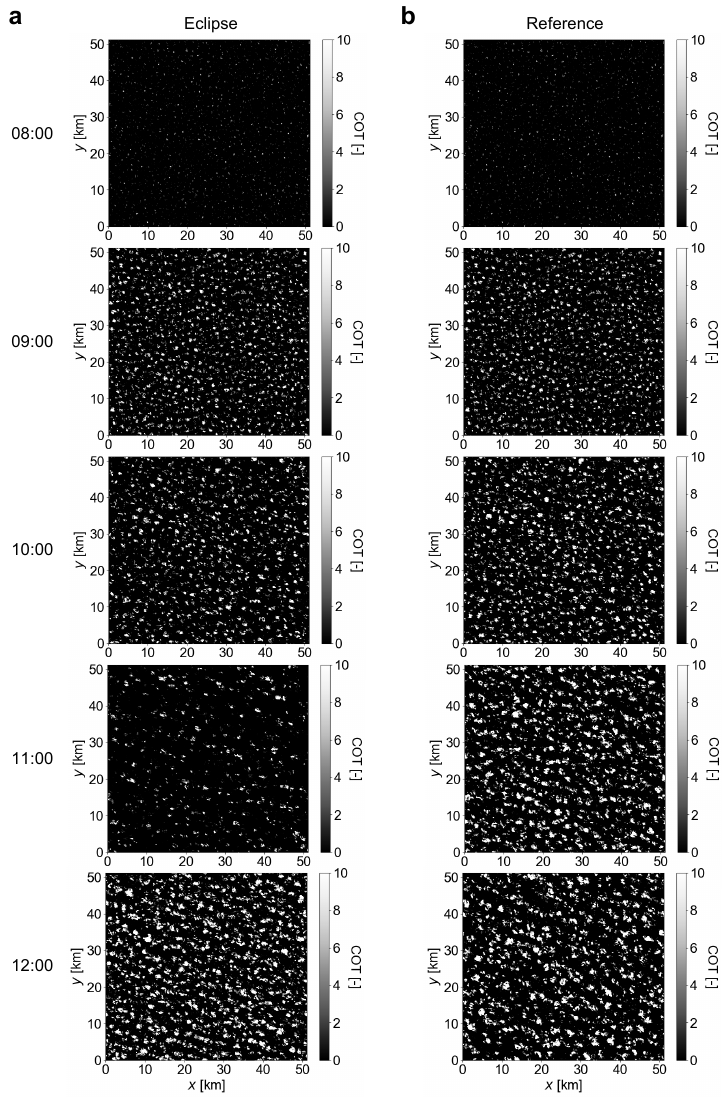}
\caption{Simulated cloud optical thickness (top view) with DALES for (a) the solar eclipse case and (b) the reference case, at 08:00, 09:00, 10:00, 11:00 and 12:00 UTC (from top to bottom).}\label{fig:cotlessnapshots}
\end{figure*}


\begin{figure*}[t!]%
\centering
\includegraphics[width=1.0\textwidth]{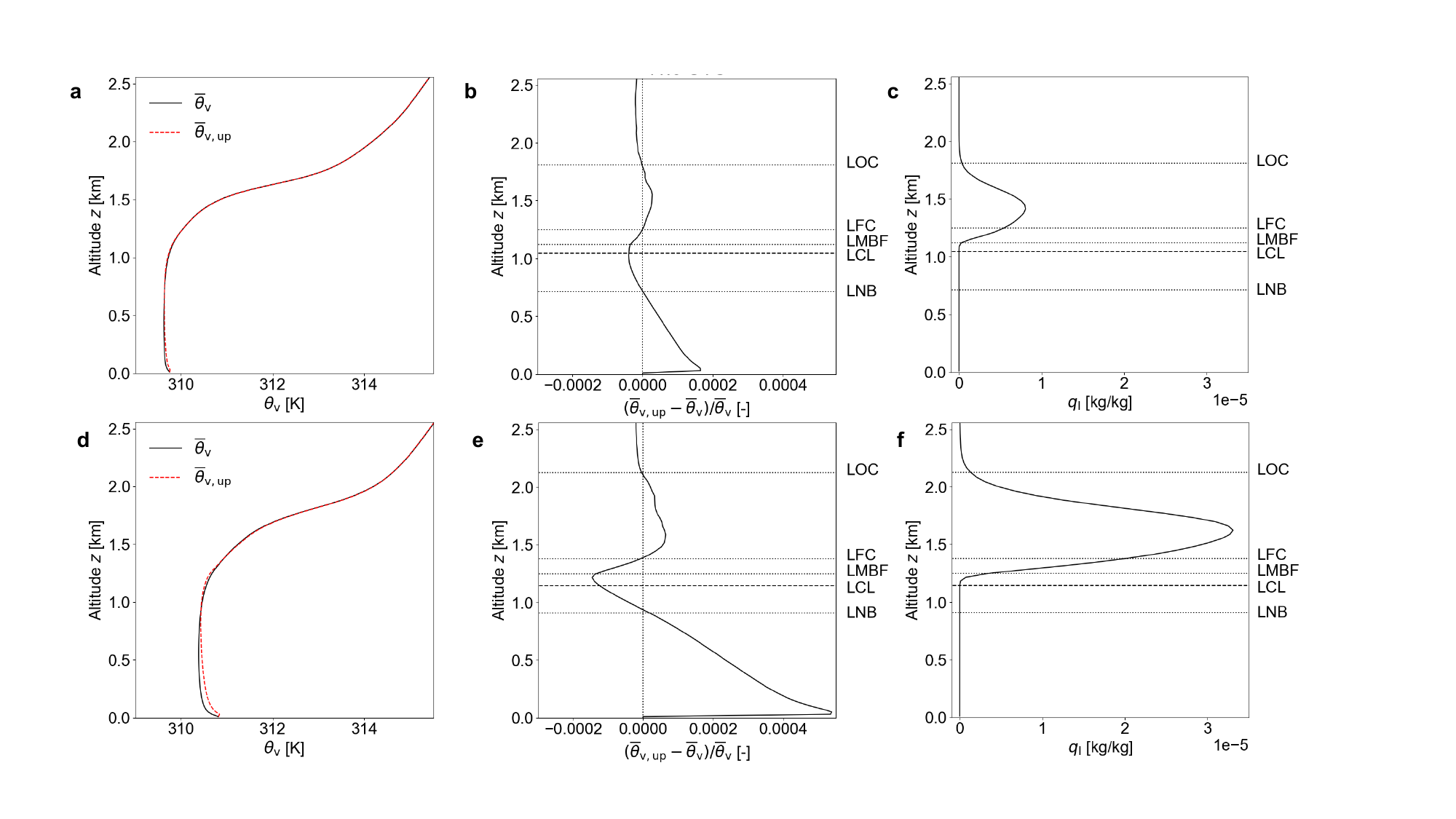}
\caption{Vertical profiles of atmospheric horizontal mean quantities simulated by DALES at 11.00 UTC. (a) The virtual potential temperature taking all columns (black solid line) and columns with updrafts only (red solid line), and (b) their normalized difference, for the solar eclipse case. (c) The liquid water specific humidity for the solar eclipse case. Similar for (d), (e) and (f) but then for the reference case. Positive values in (b) and (e) indicate positive buoyancy. In (b), (c), (e) and (f), the level of neutral buancy (LNB), lifting condensation level (LCL), level of minimum buoyancy flux (LMBF), level of free convection (LFC) and limit of convection (LOC) are indicated.}\label{fig:profiles}
\end{figure*}

\begin{figure*}[t!]%
\centering
\includegraphics[width=0.9\textwidth]{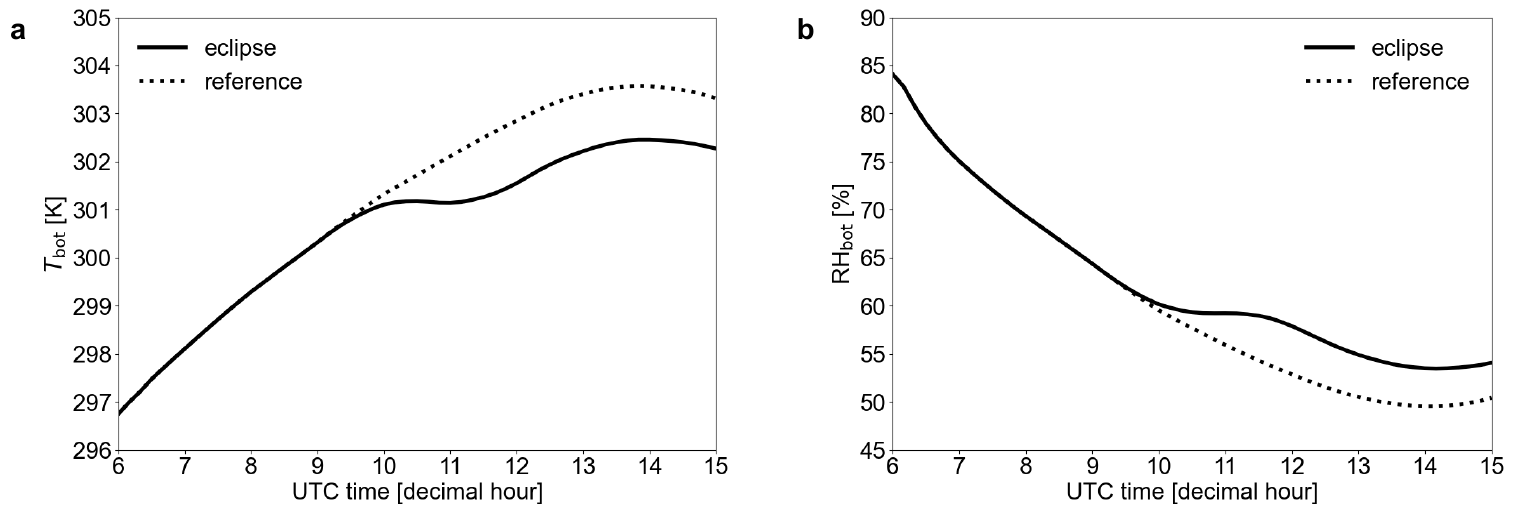}
\caption{(a) Simulated horizontal mean air temperature and (b) relative humidity by DALES of the bottom atmospheric layer (at 10 meter altitude) for the study area in the solar eclipse case (solid line) and reference case (dotted line).}\label{fig:botair}
\end{figure*}

\end{document}